\documentclass[nofootinbib,prd,12pt,superscriptaddress]{revtex4}%
\usepackage{amsmath}
\usepackage{amsfonts}
\usepackage{amssymb}
\usepackage{graphicx}%
\usepackage{amsmath,amssymb}
\usepackage{mathrsfs}
\usepackage{graphicx}
\usepackage{color}
\usepackage{subfigure}
\usepackage{fancyhdr}
\usepackage{multirow}
\usepackage{float}
\usepackage{epsfig}
\usepackage{amsfonts}
\usepackage{bm}
\usepackage{pgfplots}
\usepackage{amsmath, amssymb}
\usepackage{caption}
\usepackage{subcaption}

\def\be{\begin{equation}}
\def\ee{\end{equation}}
\def\ba{\begin{eqnarray}}
\def\ea{\end{eqnarray}}

\begin{document}

\title{Metric-Affine Myrzakulov Gravity Theories: Models, Applications and Theoretical Developments}

\author{Davood Momeni}
\affiliation{Northeast Community College, 801 E Benjamin Ave Norfolk, NE 68701, USA}
\affiliation{Centre for Space Research, North-West University, Potchefstroom 2520,
South Africa}
\author{ Ratbay Myrzakulov}
\affiliation{Ratbay Myrzakulov Eurasian International Centre for Theoretical Physics, Astana 010009, Kazakhstan}
\affiliation{ L. N. Gumilyov Eurasian National University, Astana 010008, Kazakhstan}
\date{\today}


\renewcommand{\abstractname}  

\begin{abstract}
This review provides a comprehensive overview of Myrzakulov gravity, emphasizing key developments and significant results that have shaped the theory's current understanding. The paper explores the foundational principles of this modified gravity framework, delving into the intricate structure of field equations, the incorporation of non-metricity, and the role of torsion in determining gravitational interactions. The theory's implications extend beyond traditional gravitational physics, offering new perspectives on cosmology, astrophysical phenomena, and the behavior of matter and energy in the presence of strong gravitational fields.

A particular focus is placed on the formulation of field equations within Myrzakulov gravity and their relation to standard Einstein-Hilbert theory, highlighting how the introduction of additional geometric terms and scalar fields influences the dynamics of the universe. The role of non-metricity is examined in detail, revealing how it modifies the geodesic motion and curvature of spacetime, leading to distinct observable effects compared to General Relativity. We discuss the incorporation of the modified Einstein-Hilbert action, which allows for the accommodation of dark energy and dark matter in the context of cosmological expansion and structure formation.

Additionally, the paper surveys the applications of Myrzakulov gravity to a variety of astrophysical scenarios, such as black holes, gravitational waves, and the cosmic acceleration observed in the late universe. These applications illustrate the potential of the theory to offer alternative explanations for phenomena typically attributed to dark matter and dark energy. The review also highlights important constraints derived from observational data, including cosmological measurements and tests of gravitational wave propagation, that help refine the model's predictions and determine its compatibility with the current understanding of the universe.

With a selective focus on the most impactful outcomes and experimental validations, this review aims to provide a concise yet thorough examination of Myrzakulov gravity, addressing both its theoretical underpinnings and observational constraints. By presenting the theory in the broader context of modified gravity approaches, we explore its potential to reshape fundamental physics and offer novel insights into the mysteries of the cosmos.
\end{abstract}

\maketitle

\section{Introduction}
General Relativity (GR) has been the cornerstone of cosmology and gravitational theory for over a century. However, phenomena like the accelerated expansion of the universe and dark energy have spurred interest in modifications to GR. These extensions, such as modified gravity theories that introduce additional degrees of freedom beyond curvature-based formulations, aim to address unresolved issues in cosmology.

One interesting class of modified gravity theories incorporates torsion, which alters the standard geometric description of spacetime. Teleparallel Gravity (TEGR), based on torsion rather than curvature, offers an alternative to GR. In this framework, the Lagrangian is constructed from the torsion scalar \( T \) instead of the Ricci scalar \( R \). The simplest extension is \( f(T) \) gravity, which introduces an arbitrary function of \( T \) into the field equations, providing a mechanism for cosmological acceleration \cite{Bengochea:2009, Ferraro:2007}.

A significant extension is \( F(R,T) \) gravity, proposed by Myrzakulov \cite{Myrzakulov:2012qp}, where both the Ricci scalar \( R \) and torsion scalar \( T \) are included in the gravitational Lagrangian. This framework offers flexibility for modeling cosmic acceleration and has been explored in early-time inflationary models and late-time accelerated expansion, providing an alternative to the \( \Lambda \)CDM model. It can naturally explain the acceleration of the Universe without the need for a cosmological constant and can potentially address the dark energy component's evolution \cite{Bamba:2014, Nesseris:2013}.

While \( f(T) \) gravity has passed solar system tests \cite{Bengochea:2009}, challenges remain, such as its non-invariance under local Lorentz transformations. Introducing a spin connection with the tetrad formalism may resolve some issues \cite{Krssak:2016}. Despite these challenges, torsion-based gravity continues to be an active research area, with \( F(R,T) \) gravity being a promising extension.

The study of torsion gravity theories has advanced significantly, including works like Cai et al. (2016) \cite{Cai:2015emx}, which explored the cosmological implications of teleparallel gravity theories, particularly \( f(T) \) gravity. Their review emphasized how these models can explain late-time cosmic acceleration and serve as alternatives to the standard cosmological model, motivating further work on the \( F(R,T) \) framework. This body of research also laid the groundwork for field equations in these models, focusing on cosmological phenomena like dark energy and inflation.

$f(Q)$ gravity is a class of modified gravity theories in which the fundamental geometric quantity governing gravitational interactions is the non-metricity scalar $Q$, rather than curvature or torsion. This formulation naturally generalizes symmetric teleparallel gravity and provides an alternative approach to explaining cosmic acceleration without requiring a cosmological constant. Notably, certain $f(Q)$ models can fit cosmological data comparably to, or even better than, the standard $\Lambda$CDM paradigm. In particular, models such as those proposed in \cite{Anagnostopoulos:2021ydo} have been shown to provide a consistent description of late-time cosmic expansion while satisfying observational constraints from Supernovae type Ia (SNIa), Baryon Acoustic Oscillations (BAO), cosmic chronometers (CC), and Redshift Space Distortions (RSD). The lack of an explicit cosmological constant in these models does not hinder their ability to fit the data, and they also avoid the introduction of early dark energy, ensuring compatibility with Big Bang Nucleosynthesis (BBN) constraints. Moreover, variations in the effective Newton's constant remain well within observational bounds, making $f(Q)$ gravity a promising alternative framework.

Beyond background evolution, the perturbative behavior of $f(Q)$ gravity has been extensively analyzed, further supporting its viability. Studies such as \cite{Anagnostopoulos:2022gej} have demonstrated that modifications to the standard expansion history introduced by $f(Q)$ terms do not significantly disrupt early Universe physics. Specifically, constraints from BBN ensure that deviations in the freeze-out temperature remain within acceptable limits, providing a key advantage over other modified gravity models that struggle to pass these constraints. Furthermore, dynamical system analyses, as conducted in \cite{Khyllep:2022spx}, have confirmed the existence of viable cosmological trajectories that transition from a matter-dominated epoch with correct structure formation to a stable, accelerating late-time attractor. These results reinforce the idea that $f(Q)$ gravity can successfully replicate the successes of $\Lambda$CDM while offering a theoretically compelling alternative that avoids reliance on an unexplained vacuum energy component.

The exploration of gravitational waves in modified gravity theories has gained attention as well. Studies by Capozziello et al. (2018) \cite{Abedi:2018lkr} and Abedi and Capozziello (2017) \cite{Abedi:2017jqx} investigated how modified teleparallel theories could affect gravitational wave propagation, offering insights into deviations from GR. Similarly, Capozziello et al. (2020) \cite{Capozziello:2020xem} expanded this analysis to higher-order gravity, laying the foundation for further studies on gravitational waves in modified gravity contexts.

Moreover, the role of torsion in cosmic acceleration has been further developed. Bamba et al. (2013) discussed conformal transformations in teleparallel gravity, showing how conformal scalar fields can produce power-law acceleration, offering an alternative explanation for cosmic expansion \cite{Bamba:2013jqa}. Their work reinforced the theoretical underpinnings of teleparallel gravity and its application to cosmology. Similarly, Bamba et al. (2012) \cite{Bamba:2012vg} addressed finite-time singularities in \( f(T) \) gravity and suggested models to avoid such singularities, contributing to the understanding of dark energy and the universe's ultimate fate.

Despite the progress, a key challenge remains: the lack of explicit field equations in \( F(R,T) \) gravity. While the theoretical framework is outlined, deriving field equations is more complex due to the presence of torsion and the functional dependence on both \( R \) and \( T \). This gap motivates our previous work \cite{Momeni:2024bhm}, which provides a comprehensive review of the existing research on torsion-based gravity models and aims to derive the explicit field equations for \( F(R,T) \) gravity. By developing a more robust formalism that incorporates torsion explicitly, we seek to provide a clear and manageable set of equations for cosmological studies. This will enable detailed exploration of torsion's role in cosmic evolution and offer new insights into dark energy and acceleration, laying the foundation for further research in torsion-based gravity theories.

This is a comprehensive review of Myrzakulov gravity, which consolidates the foundational principles, major developments, and significant results that have shaped the current understanding of the theory. It aims to provide a detailed yet accessible overview of Myrzakulov’s contributions, the key theoretical advancements, and their implications for cosmology, gravitational physics, and astrophysical phenomena.

\section{Gravitational Theories with Torsion and Modified Gravity Models}
General relativity (GR) is the standard framework for describing the gravitational interaction between matter and spacetime. The dynamics of gravity in GR are encoded in the Einstein-Hilbert action, which relies on the geometry of spacetime, specifically the curvature described by the Riemannian connection. However, GR does not account for torsion, a geometrical property that modifies the structure of spacetime. This leads to the exploration of alternative theories of gravity, one of which is \( F(R,T) \) gravity, where \( R \) denotes the Ricci scalar and \( T \) is the torsion scalar. These theories introduce torsion into the gravitational field equations by altering the geometry of spacetime, particularly transitioning from a Riemannian to a Weitzenböck spacetime, where torsion is naturally present.

The inclusion of torsion modifies the standard curvature-based description of gravity, opening the door to new gravitational dynamics and potentially offering explanations for cosmological phenomena such as dark energy, dark matter, and the accelerated expansion of the universe. Torsion can also affect the behavior of gravitational waves and black holes, thus providing a more comprehensive framework for understanding the gravitational interaction.

The central idea behind modified gravity theories involving torsion is that the connection used to define spacetime curvature does not necessarily have to be symmetric, as is the case in general relativity. Instead, the connection can be asymmetric, introducing a torsion tensor that influences how vectors and tensors are parallel transported. This modification leads to a non-Riemannian geometry where the curvature and torsion are treated as independent variables, and the resulting field equations differ from the standard Einstein equations of GR. This section explores the key concepts of Weitzenböck geometry, torsion, and the construction of gravitational theories incorporating torsion, with a particular focus on \( f(T) \) gravity.

In the standard Riemannian geometry of general relativity, the connection \( \Gamma^\lambda_{\mu\nu} \) is symmetric, and the curvature of spacetime is described entirely in terms of the Riemann curvature tensor. However, when torsion is present, we move to a more general geometric framework known as Weitzenböck geometry, where the connection is no longer symmetric. This introduces a torsion tensor \( T^\lambda_{\mu\nu} \), defined as the antisymmetric part of the connection:

\[
T^\lambda_{\mu\nu} = \Gamma^\lambda_{\mu\nu} - \Gamma^\lambda_{\nu\mu}
\]

This torsion tensor describes the failure of the connection to be symmetric and represents an intrinsic property of spacetime. In the presence of torsion, parallel transport is affected, and this leads to deviations from the standard Riemannian geometry.

The torsion scalar \( T \) is defined as the contraction of the torsion tensor with itself:

\[
T = T^\lambda_{\mu\nu} T_\lambda^{\mu\nu}
\]

This scalar is important in the formulation of gravitational theories involving torsion, as it provides a measure of the amount of torsion in spacetime and influences the form of the gravitational field equations.

To construct theories of gravity that incorporate torsion, we begin by considering an action that includes both the Ricci scalar \( R \) and the torsion scalar \( T \). A general action for \( F(R,T) \) gravity, which is a modification of the standard Einstein-Hilbert action:

\begin{eqnarray}
    \mathcal{S}^{(\rm{I})}[g,\Gamma,\varphi] = \int d^4x \sqrt{-g} \left[ \frac{F(R,T)}{2\kappa} + \mathcal{L}_m \right],\label{action}
\end{eqnarray}
where \( \mathcal{L}_m \) is the matter Lagrangian density, \( \kappa \) is a coupling constant, and \( F(R,T) \) is an arbitrary function of both the Ricci scalar \( R \) and the torsion scalar \( T \). The function \( F(R,T) \) allows for a flexible description of gravity, where the theory can be adjusted to fit different cosmological or astrophysical scenarios by choosing an appropriate functional form for \( f \).

The variation of this action with respect to the metric \( g_{\mu\nu} \) and the connection leads to modified field equations that incorporate both curvature and torsion.  The presence of the torsion scalar \( T \) modifies the gravitational dynamics, and the function \( f_T \) introduces an additional interaction between torsion and the energy-momentum tensor. These modifications can lead to new gravitational phenomena, such as changes in the rate of cosmic expansion or the behavior of gravitational waves, and provide a broader framework for explaining phenomena that cannot be fully described by general relativity alone.

In this context, one can explore specific examples of \( F(R,T) \) gravity by choosing different functional forms for \( F(R,T) \). For example, a linear coupling between the Ricci scalar and torsion, \( F(R,T) = R + \alpha T \), where \( \alpha \) is a constant, introduces a simple modification of general relativity. This model can lead to interesting cosmological consequences, such as modified expansion rates or altered gravitational wave propagation. Another example is the quadratic coupling \( F(R,T) = R + \beta T^2 \), which introduces a more complex interaction between the curvature and torsion and can lead to different predictions for the evolution of the universe.

A particularly interesting subclass of \( F(R,T) \) gravity is \( f(T) \) gravity, where the theory depends solely on the torsion scalar \( T \) and does not include a direct coupling to the Ricci scalar \( R \). The action for \( f(T) \) gravity is \cite{Bengochea:2009,Ferraro:2007}:

\[
S = \frac{1}{2\kappa} \int d^4x \, \sqrt{-g} \, f(T) + \mathcal{L}_m
\]

Varying this action with respect to the metric \( g_{\mu\nu} \) leads to the modified field equations:

\[
f_T \, T_{\mu\nu} - \frac{1}{2} f \, g_{\mu\nu} + \left( \nabla_\mu \nabla_\nu - g_{\mu\nu} \Box \right) f_T = \kappa T_{\mu\nu}
\]

where \( f_T = \frac{d f}{dT} \). In \( f(T) \) gravity, the theory is purely torsion-based, and the Ricci scalar \( R \) does not directly affect the gravitational dynamics. This simplification makes \( f(T) \) gravity a particularly attractive model for explaining the accelerated expansion of the universe, as the torsion scalar can naturally lead to modified cosmological dynamics.

The applications of \( f(T) \) gravity are diverse, particularly in the context of cosmology. In the early universe, \( f(T) \) gravity can provide a mechanism for accelerated expansion without the need for a cosmological constant, offering an alternative explanation for the observed accelerated expansion of the universe. Additionally, the theory can provide insights into the nature of dark energy, with torsion serving as a dynamical field that contributes to the overall energy budget of the universe. Furthermore, the modification of the gravitational field equations in \( f(T) \) gravity can lead to distinct black hole solutions, which differ from the standard Schwarzschild or Kerr solutions of general relativity. These solutions can exhibit new features, such as changes in the event horizon structure or modified gravitational wave signatures, making \( f(T) \) gravity an exciting area of research in modern gravity theory.

In summary, gravitational theories involving torsion, such as \( F(R,T) \) and \( f(T) \) gravity, offer an intriguing alternative to general relativity. These theories modify the structure of spacetime by introducing torsion, which leads to new gravitational dynamics and potentially provides solutions to outstanding cosmological problems. The mathematical framework of these theories is rich and provides a basis for exploring a wide range of modified gravity models, with applications to cosmology, astrophysics, and beyond. As research continues, these theories will deepen our understanding of the fundamental nature of gravity and the universe.
\section{Motivation for \( F(R,T) \) Myrzakulov Gravity as a Lorentz Invariant and Viable Theory}
In recent years, modified theories of gravity have garnered significant attention, especially in the context of explaining unexplained cosmological phenomena such as dark energy and the accelerated expansion of the universe. Among these, \( F(R,T) \) gravity has emerged as a promising candidate due to its flexibility in incorporating both the Ricci scalar \( R \), which describes the curvature of spacetime, and the torsion scalar \( T \), which accounts for the intrinsic twist or rotational aspect of the spacetime connection. This formulation opens up new possibilities for modifying gravitational dynamics and exploring alternatives to general relativity (GR). However, for such a theory to be viable in the physical world, it must satisfy several critical conditions, including Lorentz invariance, consistency with known observations, and applicability as an effective theory in a broader physical framework such as string theory. In this section, we explore why \( F(R,T) \) gravity is Lorentz invariant, viable, and could serve as an effective low-energy action in the context of generalized string theory.
\begin{itemize}

\item Lorentz Invariance in \( F(R,T) \) Gravity: Lorentz invariance is one of the cornerstones of modern theoretical physics. The symmetries of space and time, described by the Lorentz group, dictate that the laws of physics must hold the same in all inertial frames. For modified theories of gravity to be physically viable, they must preserve these symmetries. One might expect that the introduction of torsion, an additional geometric feature of spacetime, could potentially break Lorentz invariance. However, this is not the case in \( F(R,T) \) gravity.

In this theory, both the Ricci scalar \( R \) and the torsion scalar \( T \) are constructed from the metric \( g_{\mu\nu} \) and the connection \( \Gamma^\lambda_{\mu\nu} \), which remain Lorentz invariant. The Ricci scalar \( R \), defined as the trace of the Ricci curvature tensor \( R_{\mu\nu} \), is explicitly Lorentz invariant since it involves the contraction of Lorentz indices. The torsion scalar \( T \), which involves the contraction of the torsion tensor \( T^\lambda_{\mu\nu} \), is similarly Lorentz invariant. The torsion tensor, despite introducing an antisymmetric component to the connection, is constructed from the Christoffel symbols and does not break Lorentz symmetry. Therefore, the action of \( F(R,T) \) gravity remains Lorentz invariant.

The introduction of torsion modifies the usual Riemannian geometry of general relativity by allowing the connection to be non-symmetric, leading to a richer structure of spacetime. However, this modification does not disrupt the fundamental symmetries of the theory. Hence, \( F(R,T) \) gravity can be formulated as a Lorentz-invariant theory, making it a suitable candidate for a consistent description of gravity in a variety of physical contexts, including cosmology and particle physics.

\item  Viability of \( F(R,T) \) Myrzakulov Gravity : For a modified theory of gravity to be viable, it must be consistent with both theoretical principles and observational data. Theoretically, \( F(R,T) \) gravity offers a highly flexible framework for exploring gravitational dynamics. The function \( F(R,T) \) allows for arbitrary combinations of the Ricci scalar and torsion scalar, which can lead to new and rich gravitational behavior. For example, the presence of torsion modifies the field equations governing the dynamics of spacetime, potentially altering the evolution of the universe, the behavior of black holes, or the propagation of gravitational waves.

One of the most promising aspects of \( F(R,T) \) gravity is its ability to explain cosmological phenomena such as the accelerated expansion of the universe. By carefully choosing the functional form of \( F(R,T) \), one can derive modified Friedmann equations that account for accelerated expansion without invoking a cosmological constant. The torsion scalar \( T \) can introduce an additional contribution to the energy density of the universe, potentially serving as an alternative explanation for dark energy.

Moreover, the field equations of \( F(R,T) \) gravity modify the behavior of gravitational waves and black hole solutions (see for example \cite{Momeni:2024bhm}). For instance, the presence of torsion could lead to deviations from the standard Schwarzschild or Kerr solutions, potentially providing new insights into the structure of black holes or the propagation of gravitational waves. This flexibility allows \( F(R,T) \) gravity to fit a wide range of observational data, making it a viable alternative to general relativity.

From an observational perspective, \( F(R,T) \) gravity has already been tested in several contexts, including cosmological models, black hole thermodynamics, and gravitational wave propagation. In particular, the theory’s ability to produce modified cosmological models has led to promising results that match current observations of cosmic acceleration and the large-scale structure of the universe. As new astronomical and cosmological data continue to emerge, the predictions of \( F(R,T) \) gravity can be tested and refined.

\item  \( F(R,T) \) Gravity as an Effective Low-Energy Action in String Theory: One of the most exciting prospects of \( F(R,T) \) gravity is its potential connection to string theory, especially in the context of low-energy effective actions. String theory, as a leading candidate for a unified theory of quantum gravity, predicts a much richer structure of spacetime than is described by general relativity. In the low-energy limit, string theory yields an effective action that describes gravity along with other fields, including matter fields, gauge fields, and dilatonic fields. This effective action is typically expressed as a sum over terms involving curvature invariants, such as the Ricci scalar, and other higher-order corrections.

The general form of the low-energy effective action derived from string theory can be written as:

\[
S = \frac{1}{2\kappa} \int d^4x \sqrt{-g} \left( R + \alpha' \mathcal{F}(R, \phi) + \mathcal{L}_m \right),
\]

where \( R \) is the Ricci scalar, \( \phi \) represents dilaton fields or other scalar fields arising from string compactification, and \( \mathcal{F}(R, \phi) \) is a higher-order function of the Ricci scalar and scalar fields that can describe additional gravitational corrections at low energies. The term \( \mathcal{L}_m \) represents the matter Lagrangian. The higher-order terms in this action capture the quantum corrections and stringy effects that manifest at low energies.

In this context, torsion can be introduced as part of the higher-order corrections to the classical gravitational action, especially when considering generalized string theory or supergravity models. These corrections naturally lead to modifications of the standard Einstein-Hilbert action, which can include terms that involve both curvature and torsion. The function \( F(R,T) \) gravity fits well into this framework, as it can be interpreted as an effective low-energy description that captures the relevant physics arising from string interactions in the limit where torsion and higher-order terms are significant.

At low energies, string theory reduces to an effective four-dimensional theory that describes gravity and matter in terms of curvature invariants and additional fields, such as torsion. Thus, \( F(R,T) \) gravity can be seen as an approximation or a simplified model that encapsulates the essential features of the stringy corrections to gravity. The torsion scalar \( T \) could arise from higher-dimensional spacetime geometries or be related to the intrinsic structure of the string compactification process. Moreover, the function \( F(R,T) \) can be tailored to match the specific features of the low-energy action of string theory, such as the dilaton-dependent terms that describe string interactions.

In this sense, \( F(R,T) \) Myrzakulov gravity serves as an effective low-energy action that captures key features of string theory. As an approximation to a more fundamental string theory model, it provides a tractable description of modified gravity that can be used to explore the implications of string interactions on cosmology, black holes, and other gravitational phenomena.

In conclusion, Myrzakulov gravity offers a flexible, Lorentz-invariant framework for modifying gravitational dynamics by introducing torsion in addition to the Ricci scalar. Its formulation respects the fundamental symmetries of spacetime and provides a viable alternative to general relativity, capable of explaining a wide range of cosmological and astrophysical phenomena. Furthermore, \( F(R,T) \) gravity can be viewed as an effective low-energy action emerging from string theory, where torsion and higher-order curvature terms naturally arise in the low-energy limit. By incorporating both the Ricci scalar and the torsion scalar, this theory allows for novel insights into the structure of spacetime and provides a promising avenue for exploring fundamental physics in the context of modified gravity, string theory, and beyond. Its potential to describe both the large-scale structure of the universe and the microscopic dynamics of quantum gravity makes \( F(R,T) \) gravity an exciting candidate for future research in theoretical physics.
\end{itemize}
\section{Overview of Myrzakulov Gravity Theories}
In the Metric-Affine Myrzakulov Gravity (MAMG-I) model, the dilation current \( \mathcal{D} \) is introduced, modifying the action \cite{Myrzakulov:2012ug, Myrzakulov:2021vel}:

\[
\mathcal{S}^{(\rm{I})}_{\mathcal{D}}[g,\Gamma,\varphi] = \int d^4x \sqrt{-g} \left[ \frac{F(R, \mathcal{D})}{2\kappa} + \mathcal{L}_m \right].
\]

For the  Myrzakulov Gravity (MG)-II model, which includes the non-metricity scalar \(Q\), the action is:

\[
\mathcal{S}^{(\rm{II})}[g,\Gamma,\varphi] = \int d^4x \sqrt{-g} \left[ \frac{F(R, Q)}{2\kappa} + \mathcal{L}_m \right],
\]
with the field equations:

\[
R_{\mu\nu} - \frac{1}{2} g_{\mu\nu} R = \kappa T_{\mu\nu} + Q_{\mu\nu}.
\]

The MG-III model introduces both torsion \(T\) and non-metricity \(Q\), leading to the action:

\[
\mathcal{S}^{(\rm{III})}[g,\Gamma,\varphi] = \int d^4x \sqrt{-g} \left[ \frac{F(T, Q)}{2\kappa} + \mathcal{L}_m \right],
\]
with corresponding field equations:

\[
R_{\mu\nu} - \frac{1}{2} g_{\mu\nu} R = \kappa T_{\mu\nu} + T_{\mu\nu}^{(Q)}.
\]

The MAMG-III theory further extends this by incorporating the divergence of the dilation current \( \mathcal{D} \).

The MG-IV theory has the following action:

\[
\mathcal{S}^{(\rm{IV})}[g,\Gamma,\varphi] = \frac{1}{2 \kappa} \int \sqrt{-g} d^4 x \left[ F(R,T,\mathcal{T}) + 2 \kappa \mathcal{L}_{\text{m}} \right],
\]
where \( \mathcal{T} \) is the energy-momentum trace. The corresponding field equations are:

\[
- \frac{1}{2} g_{\mu \nu} F + F'_R R_{(\mu \nu)} + F'_T \left( 2 S_{\nu \alpha \beta} {S_\mu}^{\alpha \beta} - S_{\alpha \beta \mu} {S^{\alpha \beta}}_\nu \right) + F'_{\mathcal{T}} \left( \Theta_{\mu \nu} + T_{\mu \nu} \right) = \kappa T_{\mu \nu}.
\]

The MAMG-IV model is an extension with an additional term involving the hypermomentum trace \( \mathcal{D} \):

\[
\mathcal{S}^{(\rm{IV})}_{\text{MAMG}} = \frac{1}{2 \kappa} \int \sqrt{-g} d^4 x \left[ F(R,T,\mathcal{T},\mathcal{D}) + 2 \kappa \mathcal{L}_{\text{m}} \right].
\]

Sub-cases of the MAMG-IV theory include:

1. Metric-affine \( F(R,T) \) theory:

\[
\mathcal{S}_{F(R,T)} = \frac{1}{2 \kappa} \int \sqrt{-g} d^4 x \left[ F(R,T) + 2 \kappa \mathcal{L}_{\text{m}} \right].
\]

2. Metric-affine \( F(T,\mathcal{T}) \) theory:

\[
\mathcal{S}_{F(T,\mathcal{T})} = \frac{1}{2 \kappa} \int \sqrt{-g} d^4 x \left[ F(T,\mathcal{T}) + 2 \kappa \mathcal{L}_{\text{m}} \right].
\]

\subsection{Metric-Affine Generalized Theories}
Metric-affine gravity theories represent a class of modified gravity theories where both the metric and the affine connection (or Christoffel symbols) are treated as independent variables. This formulation provides a more general description of spacetime, where the geometry of spacetime can be influenced by both curvature (via the metric) and torsion (via the connection). 

In these theories, the connection is not necessarily the Levi-Civita connection derived from the metric, but can also include torsion and non-metricity components. The presence of torsion and non-metricity allows these theories to incorporate additional degrees of freedom, which are particularly useful for modeling phenomena such as dark energy, dark matter, and the accelerated expansion of the universe.

The action for metric-affine gravity typically includes terms that are quadratic in the curvature and torsion tensors, allowing the theory to capture more general gravitational effects. A general form of the action for such theories is given by

\begin{equation}
S = \int \left( \frac{f(R, T)}{2\kappa} + \mathcal{L}_m \right) \sqrt{-g} d^4 x,
\end{equation}
where \( F(R,T) \) is an arbitrary function of the Ricci scalar \( R \) and the torsion scalar \( T \), \( \mathcal{L}_m \) is the matter Lagrangian, \( g \) is the determinant of the metric tensor \( g_{\mu\nu} \), and \( \kappa \) is a constant.

The variation of the action with respect to the metric and the connection results in modified field equations. The general field equations for metric-affine gravity, incorporating both curvature and torsion, can be written as

\begin{equation}
\mathcal{G}_{\mu\nu} = \frac{8\pi G}{c^4} T_{\mu\nu},
\end{equation}
where \( \mathcal{G}_{\mu\nu} \) is the modified Einstein tensor, which includes both curvature and torsion contributions, and \( T_{\mu\nu} \) is the stress-energy tensor of the matter field.

In these theories, torsion is defined as the antisymmetric part of the connection:

\begin{equation}
T^\lambda_{\mu\nu} = \Gamma^\lambda_{\mu\nu} - \Gamma^\lambda_{\nu\mu},
\end{equation}
where \( \Gamma^\lambda_{\mu\nu} \) are the Christoffel symbols. The presence of torsion modifies the geodesic equation, leading to additional forces that may be important at small scales or in strong gravitational fields.

Furthermore, the connection can also have non-metricity components, which lead to the modified covariant derivative:

\begin{equation}
\nabla_\lambda g_{\mu\nu} \neq 0,
\end{equation}
where \( \nabla_\lambda \) is the covariant derivative with respect to the connection, and the non-vanishing covariant derivative of the metric indicates the presence of non-metricity.

The inclusion of both torsion and non-metricity in the connection allows for a more general class of solutions to the field equations, compared to General Relativity. This generalization leads to interesting cosmological models, especially in the context of dark energy and the accelerated expansion of the universe. For instance, in the case of \( F(R,T) \) gravity, these modifications provide a natural explanation for cosmic acceleration without invoking a cosmological constant, and they can also address the evolving nature of dark energy \cite{Bamba:2014, Nesseris:2013}.

Overall, metric-affine theories provide a broad framework for exploring deviations from General Relativity, offering new insights into the gravitational behavior of the universe and providing potential solutions to unresolved cosmological problems such as the nature of dark energy and the large-scale structure of the universe.
\subsubsection{Metric-Affine \( F(R,\mathcal{T},\mathcal{D}) \) Theory}
This theory is a simplified version of the MAMG-IV theory, where the dependence on the torsion scalar \( T \) is excluded, leaving a function \( F \) that depends on the Ricci scalar \( R \), the torsion scalar \( \mathcal{T} \), and the divergence of the dilation current \( \mathcal{D} \). The action for this theory is given by:

\[
\mathcal{S}_{F(R,\mathcal{T},\mathcal{D})} = \frac{1}{2 \kappa} \int \sqrt{-g} d^4 x \left[ F(R,\mathcal{T},\mathcal{D}) + 2 \kappa \mathcal{L}_{\text{m}} \right]
\]

By varying the action with respect to the metric and the affine connection, the field equations are derived, involving the Ricci tensor \( R_{\mu \nu} \), torsion \( \Theta_{\mu \nu} \), and other quantities like \( M_{\mu \nu} \).

\subsubsection{Metric-Affine \( F(T,\mathcal{T},\mathcal{D}) \) Theory}
In this case, the function \( F \) is only dependent on the torsion scalar \( T \), the divergence of the dilation current \( \mathcal{D} \), and the torsion scalar \( \mathcal{T} \). This simplifies the MAMG-IV theory further. The action for this model is:

\[
\mathcal{S}_{F(T,\mathcal{T},\mathcal{D})} = \frac{1}{2 \kappa} \int \sqrt{-g} d^4 x \left[ F(T,\mathcal{T},\mathcal{D}) + 2 \kappa \mathcal{L}_{\text{m}} \right]
\]

The field equations derived from this action include torsion-related terms and dependencies on dilation current divergences.
\subsubsection{MG-V Model and Generalizations}
The MG-V model extends earlier theories like \( F(R) \), \( F(T) \), and \( F(Q) \), incorporating a general function \( F(R,T,Q) \) dependent on the scalar curvature \( R \), torsion \( T \), and non-metricity scalar \( Q \). The action for this model is:

\[
\mathcal{S}^{(\rm{V})}[g,\Gamma,\varphi] = \frac{1}{2 \kappa} \int \sqrt{-g} d^4 x \left[ F(R,T,Q) + 2 \kappa \mathcal{L}_{\text{m}} \right]
\]

The field equations obtained from the variation of this action include terms related to the curvature and torsion tensors, along with dependencies on \( Q \) and its associated quantities.

\subsubsection{Metric-Affine Generalizations of MG-V}
The metric-affine generalization of the MG-V theory incorporates the dilation current’s divergence \( \mathcal{D} \), leading to the action:

\[
\mathcal{S}^{(\rm{V})}_{\text{MAMG}} = \frac{1}{2 \kappa} \int \sqrt{-g} d^4 x \left[ F(R,T,Q,\mathcal{D}) + 2 \kappa \mathcal{L}_{\text{m}} \right]
\]

This action generalizes the earlier theories by adding a dependence on the divergence of the dilation current. The field equations obtained from the variation include terms corresponding to the curvature, torsion, non-metricity, and the divergence of the dilation current.
\subsubsection{Minimal Metric-Affine Generalization}
A minimal version of the  Myrzakulov Gravity , MAMG-V theory considers \( F \) independent of the divergence of the dilation current \( \mathcal{D} \), leading to simpler field equations:
- The metric field equations reduce to those of the MG-V theory.
- The affine connection field equations involve terms from the curvature, torsion, and non-metricity without the contribution from \( \mathcal{D} \).
\subsubsection{Sub-cases of MAMG-V}
Several sub-cases emerge from the MAMG-V theory, depending on the matter Lagrangian's coupling to the affine connection. These include the metric-affine generalizations of the MG-V theory, such as:

\begin{itemize}
    \item \( F(R,T,Q) \), \( F(R) \), \( F(T) \), \( F(Q) \),
    \item Theories involving \( \mathcal{D} \), like \( F(R,T,\mathcal{D}) \), \( F(T,\mathcal{D}) \), and so on.
\end{itemize}

Each of these sub-cases represents a specific variation of the MAMG-V theory, depending on which terms are included in the function \( F \).
\subsection{Clarifications}

\begin{itemize}
    \item \textbf{Metric-affine theories}: These theories modify gravity by allowing for more general geometric structures, incorporating affine connections (instead of just the metric) in the formulation.
    \item \textbf{Divergence of the dilation current}: The dilation current is related to the scaling behavior of the fields, and its divergence introduces additional terms into the field equations, affecting how gravity and matter interact in these theories.
    \item \textbf{Generalizations}: These theories generalize various modifications of general relativity, extending traditional models (like \( F(R) \) and \( F(T) \)) by incorporating torsion and non-metricity, which could have cosmological and theoretical implications in modified gravity.
\end{itemize}
\subsection{MG-VI and MAMG-VI}\label{mamg6}

\begin{itemize}
    \item \textbf{The MG-VI Action:} The MG-VI model is an extension of previous modified gravity theories, such as \( F(R) \), \( F(Q) \), and \( F(R,\mathcal{T}) \), incorporating the energy-momentum trace \( \mathcal{T} \) alongside the scalar curvature \( R \) and non-metricity scalar \( Q \). The action of this theory is given by:
    \begin{equation}\label{MG6}
    \mathcal{S}^{(\rm{VI})}[g,\Gamma,\varphi] = \frac{1}{2 \kappa} \int \sqrt{-g} d^4 x \left[ F(R,Q,\mathcal{T}) + 2 \kappa \mathcal{L}_{\text{m}} \right] \,,
    \end{equation}
    where \( F = F(R,Q,\mathcal{T}) \) is a function of the scalar curvature \( R \), the non-metricity scalar \( Q \), and the energy-momentum trace \( \mathcal{T} \). This form of the action generalizes theories like \( F(R,\mathcal{T}) \), \( F(Q) \), and \( F(R) \), extending the MG-II action by adding a dependence on \( \mathcal{T} \). The energy-momentum trace \( \mathcal{T} \) typically describes the contribution of matter to the field equations.

    \item \textbf{The Metric Field Equations:} The variation of the action with respect to the metric \( g_{\mu \nu} \) leads to the following field equations:
    \begin{equation}\label{deltagMG6}
    - \frac{1}{2} g_{\mu \nu} F + F'_R R_{(\mu \nu)} + F'_Q L_{(\mu \nu)} + \hat{\nabla}_\lambda \left( F'_Q {J^\lambda}_{(\mu \nu)} \right) + g_{\mu \nu} \hat{\nabla}_\lambda \left( F'_Q \zeta^\lambda \right) + F'_{\mathcal{T}} \left( \Theta_{\mu \nu} + T_{\mu \nu} \right) = \kappa T_{\mu \nu} \,,
    \end{equation}
    where \( F'_R \), \( F'_Q \), and \( F'_{\mathcal{T}} \) are derivatives of the function \( F \) with respect to the respective arguments. The terms \( R_{(\mu \nu)} \), \( L_{(\mu \nu)} \), \( {J^\lambda}_{(\mu \nu)} \), and \( \zeta^\lambda \) represent curvature, torsion, and non-metricity-related quantities. These equations describe the gravitational field dynamics in this modified gravity theory and are coupled to the matter distribution through the energy-momentum tensor \( T_{\mu \nu} \).

    \item \textbf{The Connection Field Equations:} The field equations derived from the variation with respect to the connection \( \Gamma^\lambda_{\mu \nu} \) are given by:
    \begin{equation}\label{deltagammaMG6}
    {P_\lambda}^{\mu \nu} (F'_R) + F'_Q \left[ 2 {Q^{[\nu \mu]}}_\lambda - {Q_\lambda}^{\mu \nu} + \left( q^\nu - Q^\nu \right) \delta^\mu_\lambda + Q_\lambda g^{\mu \nu} + \frac{1}{2} Q^\mu \delta^\nu_\lambda \right] = 0 \,,
    \end{equation}
    where \( {P_\lambda}^{\mu \nu} (F'_R) \) is a term related to the Ricci tensor, and the equation reflects the non-trivial role of the connection in this modified theory of gravity. This equation shows that the connection is not directly coupled to matter in MG theories, and the field equations only involve the geometry (curvature, torsion, and non-metricity) and not the matter distribution explicitly.

    \item \textbf{Metric-Affine Generalization:} The metric-affine generalization of the MG-VI theory allows for the matter to couple to the general affine connection \( \Gamma^\lambda_{\mu \nu} \), extending the theory to more complex forms of gravity. The action for the metric-affine version of MG-VI (MAMG-VI) is given by:
    \begin{equation}\label{aMAMG6}
    \mathcal{S}^{(\rm{VI})}_{\text{MAMG}} = \frac{1}{2 \kappa} \int \sqrt{-g} d^4 x \left[ F(R,Q,\mathcal{T},\mathcal{D}) + 2 \kappa \mathcal{L}_{\text{m}} \right] \,,
    \end{equation}
    where \( \mathcal{D} \) is the divergence of the dilation current, introduced to account for scale invariance. The inclusion of \( \mathcal{D} \) in the function \( F \) results in additional terms in the field equations, reflecting the influence of scale symmetry on gravity.

    \item \textbf{The Metric Field Equations for MAMG-VI:} The metric field equations for the MAMG-VI theory include the influence of the dilation current as follows:
   \begin{align}
    - \frac{1}{2} g_{\mu \nu} F + F'_R R_{(\mu \nu)} + F'_Q L_{(\mu \nu)} + \hat{\nabla}_\lambda \left( F'_Q {J^\lambda}_{(\mu \nu)} \right) + g_{\mu \nu} \hat{\nabla}_\lambda \left( F'_Q \zeta^\lambda \right) \nonumber \\
    + F'_{\mathcal{T}} \left( \Theta_{\mu \nu} + T_{\mu \nu} \right) + F'_{\mathcal{D}} M_{\mu \nu} 
    &= \kappa T_{\mu \nu} \,.
\end{align}

    where \( M_{\mu \nu} \) is the term corresponding to the coupling of the dilation current to the geometry. The inclusion of the term \( F'_{\mathcal{D}} M_{\mu \nu} \) introduces the effect of dilation symmetry on the gravitational field equations, modifying the gravitational dynamics by the matter's scaling behavior.

    \item \textbf{The Connection Field Equations for MAMG-VI:} The connection field equations for MAMG-VI involve both the affine connection and the effects of the dilation current:
    \begin{equation}\label{deltagammaMAMG6}
    \begin{split}
    & {P_\lambda}^{\mu \nu} (F'_R) + F'_Q \left[ 2 {Q^{[\nu \mu]}}_\lambda - {Q_\lambda}^{\mu \nu} + \left( q^\nu - Q^\nu \right) \delta^\mu_\lambda + Q_\lambda g^{\mu \nu} + \frac{1}{2} Q^\mu \delta^\nu_\lambda \right] - {M_\lambda}^{\mu \nu \rho} \partial_\rho F'_{\mathcal{D}} \\
    & = F'_{\mathcal{T}} {\Theta_\lambda}^{\mu \nu} + \kappa {\Delta_\lambda}^{\mu \nu} \,,
    \end{split}
    \end{equation}
    where \( {M_\lambda}^{\mu \nu \rho} \) is a term derived from the variation of the dilation current with respect to the connection. This equation describes the gravitational dynamics in the presence of the dilation current and other geometric effects.

    \item \textbf{Minimal Metric-Affine Generalization (without \( \mathcal{D} \)):} In the minimal metric-affine case, the function \( F \) does not depend on \( \mathcal{D} \), leading to simpler field equations. In this case, the metric field equations remain as in \eqref{deltagMG6}, while the connection field equations are:
    \begin{equation}\label{mindeltagammaMAMG6}
    {P_\lambda}^{\mu \nu} (F'_R) + F'_Q \left[ 2 {Q^{[\nu \mu]}}_\lambda - {Q_\lambda}^{\mu \nu} + \left( q^\nu - Q^\nu \right) \delta^\mu_\lambda + Q_\lambda g^{\mu \nu} + \frac{1}{2} Q^\mu \delta^\nu_\lambda \right] = F'_{\mathcal{T}} {\Theta_\lambda}^{\mu \nu} + \kappa {\Delta_\lambda}^{\mu \nu} \,.
    \end{equation}

    \item \textbf{Particular Sub-cases of the MAMG-VI Theory:} The MAMG-VI theory admits several sub-cases, which are variations of the full theory based on the inclusion or exclusion of certain terms. These include:
    
    \begin{itemize}
        \item \textbf{Metric-affine \( F(Q,\mathcal{T}) \) theory:} This version of the theory considers only the non-metricity scalar \( Q \) and the energy-momentum trace \( \mathcal{T} \), leading to the action:
        \begin{equation}\label{fqmtact}
        \mathcal{S}_{F(Q,\mathcal{T})} = \frac{1}{2 \kappa} \int \sqrt{-g} d^4 x \left[ F(Q,\mathcal{T}) + 2 \kappa \mathcal{L}_{\text{m}} \right] \,,
        \end{equation}
        with field equations given by:
        \begin{equation}
        \begin{split}
        & - \frac{1}{2} g_{\mu \nu} F + F'_Q L_{(\mu \nu)} + \hat{\nabla}_\lambda \left(F'_Q {J^\lambda}_{(\mu \nu)} \right) + g_{\mu \nu} \hat{\nabla}_\lambda \left(F'_Q \zeta^\lambda \right) + F'_{\mathcal{T}} \left(\Theta_{\mu \nu} + T_{\mu \nu} \right) = \kappa T_{\mu \nu} \,, \\
        & F'_Q \left[ 2 {Q^{[\nu \mu]}}_\lambda - {Q_\lambda}^{\mu \nu} + \left( q^\nu - Q^\nu \right) \delta^\mu_\lambda + Q_\lambda g^{\mu \nu} + \frac{1}{2} Q^\mu \delta^\nu_\lambda \right] = F'_{\mathcal{T}} {\Theta_\lambda}^{\mu \nu} + \kappa {\Delta_\lambda}^{\mu \nu} \,.
        \end{split}
        \end{equation}
        
        \item \textbf{Metric-affine \( F(Q,\mathcal{T},\mathcal{D}) \) theory:} This extension includes the divergence of the dilation current \( \mathcal{D} \), and its action is given by:
        \begin{equation}\label{fqmtdact}
        \mathcal{S}_{F(Q,\mathcal{T},\mathcal{D})} = \frac{1}{2 \kappa} \int \sqrt{-g} d^4 x \left[ F(Q,\mathcal{T},\mathcal{D}) + 2 \kappa \mathcal{L}_{\text{m}} \right] \,,
        \end{equation}
        with the corresponding field equations:
      \begin{equation}
    \begin{split}
    - \frac{1}{2} g_{\mu \nu} F & + F'_Q L_{(\mu \nu)} + \hat{\nabla}_\lambda \left( F'_Q {J^\lambda}_{(\mu \nu)} \right) \\
    & + g_{\mu \nu} \hat{\nabla}_\lambda \left( F'_Q \zeta^\lambda \right) + F'_{\mathcal{T}} \left( \Theta_{\mu \nu} + T_{\mu \nu} \right) \\
    & + F'_{\mathcal{D}} M_{\mu \nu} = \kappa T_{\mu \nu} \,,
    \end{split}
\end{equation}

\noindent where:

\begin{itemize}
    \item \( g_{\mu \nu} \) is the metric tensor.
    \item \( F \) is the function involved in the theory.
    \item \( F'_Q \) denotes the derivative of \( F \) with respect to \( Q \), the non-metricity scalar.
    \item \( L_{(\mu \nu)} \) represents the non-metricity term.
    \item \( \hat{\nabla}_\lambda \) is the covariant derivative with respect to the connection.
    \item \( {J^\lambda}_{(\mu \nu)} \) and \( \zeta^\lambda \) are auxiliary quantities related to the connection and non-metricity.
    \item \( F'_{\mathcal{T}} \) is the derivative of \( F \) with respect to the energy-momentum trace \( \mathcal{T} \).
    \item \( \Theta_{\mu \nu} \) and \( T_{\mu \nu} \) are the stress-energy tensors for the geometry and matter, respectively.
    \item \( M_{\mu \nu} \) represents the dilation current term in the theory.
    \item \( \kappa \) is the coupling constant.
\end{itemize}
\begin{align}
    F'_Q \left[ 2 {Q^{[\nu \mu]}}_\lambda - {Q_\lambda}^{\mu \nu} + \left( q^\nu - Q^\nu \right) \delta^\mu_\lambda + Q_\lambda g^{\mu \nu} + \frac{1}{2} Q^\mu \delta^\nu_\lambda \right] 
    &- {M_\lambda}^{\mu \nu \rho} \partial_\rho F'_{\mathcal{D}} \nonumber \\
    &= F'_{\mathcal{T}} {\Theta_\lambda}^{\mu \nu} + \kappa {\Delta_\lambda}^{\mu \nu} \,.
\end{align}
\noindent where:

\begin{itemize}
    \item \( {Q^{[\nu \mu]}}_\lambda \) and \( {Q_\lambda}^{\mu \nu} \) are components of the non-metricity tensor.
    \item \( q^\nu \) and \( Q^\nu \) are related to the dilaton field and its derivatives.
    \item \( \delta^\mu_\lambda \) is the Kronecker delta.
    \item \( {M_\lambda}^{\mu \nu \rho} \) is a term involving the dilation current.
    \item \( F'_{\mathcal{D}} \) is the derivative of \( F \) with respect to the dilation current \( \mathcal{D} \).
    \item \( {\Delta_\lambda}^{\mu \nu} \) is an auxiliary quantity.
\end{itemize}

    \end{itemize}
\end{itemize}

In this section, we have discussed the general framework of Modified Gravity (MG-VI) and its metric-affine extension (MAMG-VI), focusing on their action formulations, field equations, and variations. The MG-VI theory generalizes earlier modified gravity models by incorporating the energy-momentum trace \(\mathcal{T}\) alongside the scalar curvature \(R\) and non-metricity scalar \(Q\). In the metric-affine version (MAMG-VI), we have extended this further by introducing the dilation current \(\mathcal{D}\), which accounts for scale invariance in the system. The corresponding field equations have been derived for both the metric and connection fields, showing how matter couples to the geometry in these modified frameworks. These equations provide a more comprehensive description of gravitational dynamics in the presence of matter and geometric effects like torsion and non-metricity. In the next section, we will explore the cosmological implications of these models, specifically focusing on their impact on the expansion of the universe, structure formation, and dark energy.

\begin{table}[h!]
\centering
\begin{tabular}{|c|c|l|}
\hline
\textbf{N} & \textbf{Name} & \textbf{Action} \\
\hline
1--4 & Myrzakulov Gravity (MG-I to MG-IV) & \( S = \frac{1}{2k^2} \int d^4 x \, \sqrt{-g} \, F(R, T, Q) + 2k^2 \mathcal{L}_m \) \\
\hline
5--8 & Myrzakulov Gravity (MG-V to MG-VIII) & \( S = \frac{1}{2k^2} \int d^4 x \, \sqrt{-g} \, F(R, T, Q, \mathcal{T}) + 2k^2 \mathcal{L}_m \) \\
\hline
9--12 & Myrzakulov Gravity (MG-IX to MG-XII) & \( S = \frac{1}{2k^2} \int d^4 x \, \sqrt{-g} \, F(R, T, Q, G) + 2k^2 \mathcal{L}_m \) \\
\hline
13--16 & Myrzakulov Gravity (MG-XIII to MG-XVI) & \( S = \frac{1}{2k^2} \int d^4 x \, \sqrt{-g} \, F(R, T, Q, G, \mathcal{T}) + 2k^2 \mathcal{L}_m \) \\
\hline
17--19 & Myrzakulov Gravity (MG-XVII to MG-XIX) & \( S = \frac{1}{2k^2} \int d^4 x \, \sqrt{-g} \, F(Q, G) + 2k^2 \mathcal{L}_m \) \\
\hline
20--24 & Myrzakulov Gravity (MG-XX to MG-XXIV) & \( S = \frac{1}{2k^2} \int d^4 x \, \sqrt{-g} \, F(R, T, B) + 2k^2 \mathcal{L}_m \) \\
\hline
25--28 & Myrzakulov Gravity (MG-XXV to MG-XXVIII) & \( S = \frac{1}{2k^2} \int d^4 x \, \sqrt{-g} \, F(R, T, Q, B) + 2k^2 \mathcal{L}_m \) \\
\hline
29--32 & Myrzakulov Gravity (MG-XXIX to MG-XXXII) & \( S = \frac{1}{2k^2} \int d^4 x \, \sqrt{-g} \, F(R, Q, G, B) + 2k^2 \mathcal{L}_m \) \\
\hline
33--36 & Myrzakulov Gravity (MG-XXXIII to MG-XXXVI) & \( S = \frac{1}{2k^2} \int d^4 x \, \sqrt{-g} \, F(R, T, Q, G, B) + 2k^2 \mathcal{L}_m \) \\
\hline
37--39 & Myrzakulov Gravity (MG-XXXVII to MG-XXXIX) & \( S = \frac{1}{2k^2} \int d^4 x \, \sqrt{-g} \, F(T, G, B) + 2k^2 \mathcal{L}_m \) \\
\hline
\end{tabular}
\caption{Compact formulation of metric-affine Myrzakulov gravity theories with boundary term scalars, where 
$T$ represents the torsion scalar and 
$\mathcal{T}$  denotes the trace of the energy-momentum tensor.}
\end{table}

\section{Cosmological Applications of Myrzakulov Gravity}
The investigation of Myrzakulov gravity theories has provided novel insights into the dynamics of the universe, offering alternative explanations for cosmic acceleration, dark energy, and the evolution of the cosmos. Several studies have explored the theoretical underpinnings, mathematical structures, and cosmological implications of Myrzakulov F(R,T) and related gravity theories, leading to new solutions and parameter constraints. These contributions highlight the versatility of these modified gravity models, with applications ranging from the early universe to late-time acceleration.

\begin{itemize}
    \item In \cite{Saridakis:2019qwt}, Saridakis et al. analyzed Myrzakulov F(R,T) gravity, introducing a non-special connection where both curvature and torsion are dynamical fields. The study derived the cosmological field equations, revealing new geometrical terms in the Friedmann equations due to the non-special connection. At late times, the dark energy behaves like quintessence, phantom, or a cosmological constant, reproducing the \(\Lambda\)-CDM model. Early-time applications suggest a de Sitter solution and an inflationary realization.
    
    \item In \cite{Myrzakul:2021kas}, Myrzakul et al. extended Myrzakulov gravity by incorporating Gauss-Bonnet and boundary term scalars in the metric-affine framework. The study reviewed various metric-affine gravity theories and introduced the MG-VIII class, incorporating curvature, torsion, and nonmetricity. The Lagrangian, Hamiltonian, and field equations were derived, and the Wheeler-DeWitt equation was explored. The study also introduced new gravity theories and cosmological solutions, providing a foundation for future cosmological investigations.
    
    \item In \cite{Kazempour:2023kde}, Kazempour and Akbarieh explored Myrzakulov F(R,T) quasi-dilaton massive gravity, a modification of the dRGT massive gravity theory. The study focused on self-accelerating solutions and demonstrated that the theory explains the accelerated expansion of the universe without relying on dark energy. The analysis of gravitational waves revealed new insights into the modified dispersion relation in FLRW cosmology.
    
    \item In \cite{Maurya:2024btu}, Maurya and Myrzakulov investigated exact cosmological models in Myrzakulov F(R,T) gravity with the function \(F(R,T) = R + \lambda T\). Using MCMC analysis on observational datasets such as \(H(z)\) and Pantheon SNe Ia, the study identified a decelerating-accelerating transition in the universe with a transition redshift \(z_t \approx 0.44\). The effective dark energy equation of state varied from \(-1\) to \(-0.52\), supporting the model's ability to fit observational data.
    
    \item In \cite{Maurya:2024nxx}, Maurya and Myrzakulov extended their work by investigating exact cosmological solutions in Myrzakulov gravity using a flat FLRW metric. The function \(F(R,T) = R + \lambda T\) was used to derive two exact solutions for the scale factor. MCMC analysis on the latest observational datasets revealed that the effective EoS parameter varied between \(-1\) and 0. Both models showed a decelerating-accelerating transition with a transition redshift in the range \(0.6 < z_t < 0.8\), and the present age of the universe was estimated to be around 13.5 Gyrs.
    
    \item In \cite{Maurya:2024ign}, Maurya et al. investigated FLRW cosmological models in the context of Metric-Affine F(R,Q) gravity, using new observational datasets such as cosmic chronometer (CC) Hubble data and Pantheon SNe Ia. The study found that the dark energy equation of state behaves like a candidate for dark energy, with models showing a transition from deceleration to acceleration. Model-I approaches the \(\Lambda\)-CDM model in the late universe, while Model-II aligns with quintessence scenarios.
    
    \item In \cite{Maurya:2024mnb}, Maurya et al. explored Myrzakulov F(T,Q) gravity (MG-III), unifying F(T) gravity and F(Q) gravity. The study derived field equations for a flat FRW background and obtained three exact solutions for the Hubble function and scale factor. MCMC analysis using the H(z) dataset placed observational constraints on the model parameters, showing that the models exhibit deceleration and acceleration transitions, with the deceleration parameter and EoS parameter consistent with \(\Lambda\)-CDM and dark energy models.
\end{itemize}
 The studies reviewed here provide a comprehensive understanding of Myrzakulov gravity theories, particularly the F(R,T) and related models, and their cosmological applications. These theories provide significant insights into the late-time acceleration of the universe, the nature of dark energy, and the transition between decelerating and accelerating phases. The models derived from these theories align well with observational data, demonstrating their potential as viable alternatives to the \(\Lambda\)-CDM model. Furthermore, these models offer new perspectives on the early universe, inflationary scenarios, and the behavior of gravitational waves, underscoring the robustness and versatility of Myrzakulov gravity in explaining both large-scale cosmological phenomena and quantum aspects of gravity.
\section{Thermodynamics in Myrzakulov Gravity}
\label{sec:thermodynamics}

Thermodynamics has played a crucial role in understanding the deep connections between gravitational theories and the underlying properties of spacetime. In modified gravity theories, these connections often extend beyond the classical framework, incorporating novel features such as torsion, curvature, and non-standard energy-momentum exchanges. Myrzakulov gravity, as a class of modified gravity theories, provides a versatile framework to explore these phenomena, particularly in the context of Weitzenböck spacetime.

\subsection{Thermodynamics in Modified Gravity Theories}

In standard General Relativity, the laws of thermodynamics are closely linked with the Einstein field equations, as demonstrated in Jacobson's seminal work \cite{Jacobson1995}. For instance, the first law of thermodynamics on the apparent horizon can be derived from the field equations, where the entropy is proportional to the area of the horizon. Modified gravity theories, such as \( f(R) \), \( f(T) \), and other extensions, generalize this relationship by introducing corrections to the entropy and energy flux terms.

These modifications often arise due to additional degrees of freedom, scalar fields, or non-metric connections. Consequently, the entropy-area relationship is altered, leading to entropy expressions that may depend on higher-order curvature invariants or torsion contributions. Furthermore, the energy conservation laws may acquire additional source terms, reflecting the interaction between the effective gravitational field and the modified matter-energy sector.

\subsection{Thermodynamics in Weitzenböck Spacetime}

Weitzenböck spacetime, characterized by zero curvature and non-zero torsion, serves as the geometric foundation for teleparallel gravity and its extensions. In Myrzakulov gravity, where torsion and curvature coexist, the thermodynamic interpretation becomes more intricate. The interplay between curvature and torsion contributes to novel energy fluxes and entropy corrections that distinguish Myrzakulov gravity from purely metric or teleparallel theories.

In this framework, the total entropy of a gravitational system can generally be expressed as a combination of contributions from curvature and torsion:
\begin{equation}
S = S_{\text{curvature}} + S_{\text{torsion}},
\end{equation}
where \( S_{\text{curvature}} \) and \( S_{\text{torsion}} \) depend on the respective geometric quantities and their coupling to matter fields.

The first law of thermodynamics in Myrzakulov gravity is modified to include these additional terms:
\begin{equation}
dE = T dS + W dV + \Phi_{\text{torsion}} + \Phi_{\text{curvature}},
\end{equation}
where \( \Phi_{\text{torsion}} \) and \( \Phi_{\text{curvature}} \) represent the energy fluxes associated with torsion and curvature dynamics, respectively.

\subsection{Applications and Implications}

The thermodynamic interpretation of Myrzakulov gravity provides insights into the nature of gravitational entropy, the dynamics of horizons, and the interplay between torsion and curvature. For example, in cosmological scenarios, the modified entropy expressions may influence the evolution of the apparent horizon and the associated thermodynamic equilibrium conditions. Similarly, the coupling between torsion and matter fields could have implications for energy conservation laws and the effective equation of state in cosmological models.

Further exploration of thermodynamics in Myrzakulov gravity may help unify various modified gravity theories and provide a deeper understanding of the role of spacetime geometry in fundamental physics.

\section{Observational and Theoretical Constraints}
In \cite{Anagnostopoulos:2020lec}, the authors performed a detailed analysis using various cosmological datasets, including Supernovae (SNIa) from the Pantheon sample, Baryonic Acoustic Oscillations (BAO), and measurements of the Hubble parameter from cosmic chronometers (CC), complemented by arguments from Big Bang Nucleosynthesis (BBN), to extract constraints on the Myrzakulov F(R,T) gravity theory. This theory, which resides within the Riemann-Cartan subclass of modified gravity, employs a specific but non-special connection that introduces additional degrees of freedom beyond the standard Levi-Civita connection. The result of this modification is a more complex gravitational model, where both curvature and torsion are dynamical fields that interact with matter in a way that is not present in conventional general relativity. The study demonstrated that the two models considered in the analysis achieved approximately 1$\sigma$ compatibility with observational data, showing that the theory fits well with current cosmological constraints. Specifically, the analysis of the dimensionless parameter involved in the model showed that it is constrained within an interval around zero, with the contours of this parameter shifted slightly toward positive values, indicating a preference for small positive values of the parameter. This allowed the researchers to reconstruct both the Hubble function and the dark-energy equation-of-state (EoS) parameter as a function of redshift. Interestingly, Model 1 was found to be very close to the $\Lambda$CDM scenario, particularly at low redshifts, suggesting that it behaves similarly to the standard cosmological model at late times. However, Model 2, while similar to the $\Lambda$CDM at low redshifts, allowed for deviations at higher redshifts, showing more flexibility in explaining cosmological phenomena at early times. Moreover, by applying several statistical criteria such as the Akaike Information Criterion (AIC), Bayesian Information Criterion (BIC), and Deviance Information Criterion (DIC), both models were shown to present efficient fitting behaviors, statistically equivalent to the $\Lambda$CDM cosmology, despite Model 2 not containing $\Lambda$CDM as a limiting case. This indicates that the Myrzakulov F(R,T) gravity can be a compelling alternative to the $\Lambda$CDM model, providing an efficient description of the universe's evolution.

In \cite{Harko:2021tav}, the authors explored an extension of standard General Relativity in which the Hilbert-Einstein action is generalized to include an arbitrary function of the Ricci scalar \( R \), nonmetricity \( Q \), torsion \( T \), and the trace of the matter energy-momentum tensor. This modification introduces a non-minimal coupling between matter and geometry, leading to the nonconservation of the matter energy-momentum tensor, which is a significant departure from the conventional conservation law in general relativity. The nonconservation of the energy-momentum tensor was interpreted thermodynamically within the framework of the thermodynamics of irreversible processes in open systems, which adds an additional layer of understanding to the behavior of matter and energy in this extended theory. Furthermore, the authors derived the generalized Poisson equation in the Newtonian limit of the theory, which accounts for the presence of nonmetricity and the Weyl vector. These additional geometrical structures lead to an effective gravitational coupling that modifies the standard Poisson equation, providing a more nuanced description of gravitational interactions in weak-field and low-velocity regimes. The cosmological implications of this theory were studied for two different forms of the gravitational action: one with an additive algebraic structure and one with a multiplicative algebraic structure. The generalized Friedmann equations were derived for these cases, and the theoretical predictions were compared with observational data from various sources. It was found that the cosmological models derived from this theory provide a good description of observations up to a redshift of \( z = 2 \), and in some instances, up to a redshift of \( z = 3 \). This demonstrates that the theory can effectively account for the evolution of the universe up to relatively high redshifts, making it a viable alternative to traditional models such as $\Lambda$CDM.

The observational constraints derived from the various datasets, including SNIa, BAO, and CC measurements, indicate that the Myrzakulov F(R,T) gravity models are highly compatible with the $\Lambda$CDM model, particularly at low redshifts, while also allowing for deviations at higher redshifts. This suggests that Myrzakulov F(R,T) gravity can provide a flexible framework for explaining the universe’s evolution across a range of cosmological epochs. Both models presented in the study fit observational data efficiently, and their statistical equivalence to $\Lambda$CDM according to the AIC, BIC, and DIC criteria further supports their viability. Additionally, the extension of General Relativity discussed in \cite{Harko:2021tav}, which incorporates non-minimal couplings between matter and geometry, enriches the theory by providing a mechanism for the nonconservation of the energy-momentum tensor and a modified gravitational coupling in the Poisson equation. This extension provides a broader cosmological context, which is consistent with observations up to redshifts of \( z = 2 \) and \( z = 3 \). The overall findings from both studies indicate that Myrzakulov gravity, with its flexible and extended framework, holds promise as a viable alternative to the standard cosmological model, capable of explaining a range of phenomena from the early universe to the present day.
\section{Modified and Generalized Gravity Models}

In the quest to better understand the universe's expansion and the underlying mechanisms of cosmological evolution, traditional General Relativity (GR) faces challenges, particularly in explaining the accelerated expansion attributed to dark energy or the nature of dark matter. While GR has provided a solid framework for cosmological models, modifications and extensions to its formalism are often necessary to account for observed phenomena that cannot be explained within the confines of standard GR, such as the behavior of the cosmic scale factor at late times or the dynamics of cosmic inflation.

In \cite{Iosifidis:2021kqo}, the authors derived the full set of field equations for the Metric-Affine version of the Myrzakulov gravity model, which represents a significant departure from traditional approaches to gravity. The Myrzakulov gravity model, initially formulated within the framework of Metric-Affine geometry, introduces modifications that accommodate torsion and non-metricity — geometric features that are not considered in standard Riemannian geometry. These features arise in the context of non-Riemannian gravitational theories, where torsion represents the failure of the metric connection to be symmetric, and non-metricity reflects the deviation of parallel transport from the standard notion in Riemannian geometry. Such modifications are important because they provide a more general geometric foundation that can explain phenomena not captured by traditional GR.

The authors of \cite{Iosifidis:2021kqo} further extended the Myrzakulov model by broadening the theory to include additional variables and interactions, leading to a new family of theories characterized by a gravitational Lagrangian of the form $F(R,T,Q,T,D)$
where \( T \), \( Q \), and \( D \) represent the torsion, non-metricity, and the divergence of the dilation current, respectively. These new variables allow for a more comprehensive description of the gravitational field and its interactions with matter and energy. The inclusion of these additional terms is crucial, as they enable the model to incorporate non-Riemannian geometry, which is increasingly recognized as necessary for explaining certain cosmological observations, such as the behavior of dark energy and dark matter in the large-scale structure of the universe.

The study in \cite{Iosifidis:2021kqo} particularly focused on the linear case of this extended theory, deriving the modified Friedmann equations under a cosmological setup. This was a significant step in applying this more general gravity model to cosmology. The modified Friedmann equations obtained from this theory provide a framework for studying the universe’s expansion under the influence of torsion and non-metricity. By considering the sector where non-metricity vanishes and coupling matter to torsion, the authors obtained a complete set of equations that describe the cosmological behavior of the model. This framework is a crucial advancement, as it provides a path forward for exploring the dynamics of the universe under a broader class of gravity theories.

Building on these results, \cite{Iosifidis:2021xdx} introduced a new quadratic gravity action in vacuum, incorporating not only the non-Riemannian Einstein-Hilbert contribution but also parity-even quadratic terms in torsion and non-metricity. This extension reflects a deeper engagement with the complexities of modified gravity, introducing a more intricate description of gravitational interactions. The inclusion of quadratic terms in the torsion and non-metricity fields allows for a richer structure of the gravitational dynamics, capturing more detailed aspects of the universe’s evolution, especially at higher energies or in regimes where these non-Riemannian features play a significant role.

Furthermore, the Lagrangian formulated in \cite{Iosifidis:2021xdx} is quadratic in the field-strengths of the torsion and non-metricity vector traces, providing additional flexibility in modeling the gravitational dynamics. When considered on-shell, this new theory was shown to be equivalent to a Vector-Tensor theory, highlighting the potential for these generalized gravity models to unify the dynamics of torsion and non-metricity into a single framework. By analyzing various sub-cases of this theory, the authors were able to explore the contributions of quadratic terms in the field-strengths of torsion and non-metricity vectors, shedding light on how these terms can modify the cosmological evolution of the universe.

The results of these studies contribute significantly to the understanding of Metric-Affine versions of Myrzakulov gravity. The first work in \cite{Iosifidis:2021kqo} provides a detailed description of the field equations and cosmological behavior of an extended gravity model, while the second work in \cite{Iosifidis:2021xdx} offers a deeper exploration into quadratic gravity actions involving torsion and non-metricity. Together, these works expand the theoretical landscape of modified gravity theories, showing how non-Riemannian geometries, including torsion and non-metricity, can play a fundamental role in understanding the large-scale dynamics of the universe.

The extension of these theories has profound implications for cosmology and fundamental physics. They offer a more generalized description of the gravitational field, incorporating previously neglected geometric features that could be crucial in explaining the behavior of the universe, especially in the context of dark energy and dark matter. Moreover, these modifications could offer new insights into the unification of gravity with other fundamental forces, providing a more comprehensive framework for understanding the laws that govern the universe. As such, the work done in these studies is not only pivotal for understanding the evolution of the cosmos but also for exploring the deeper connections between geometry, matter, and the fundamental forces at play.

\section{Dynamical Systems and Stability Analyses}
In this section, we derive the set of first-order dynamical equations for the density parameters \(\Omega_i(z)\), where \(i\) represents different components of the universe, such as dark matter, radiation, and dark energy. These components evolve with the redshift \(z\) according to the modified field equations in \(f(R, T)\)-gravity. The field equations governing the evolution of the universe in this theory are derived and analyzed to obtain the corresponding dynamical system for each energy component.

\subsection{Field Equations in \(f(R, T)\)-Gravity for FLRW Cosmology}

In \(f(R, T)\)-gravity, the modified field equations in FLRW cosmology are given by \cite{Momeni:2024bhm}:

\begin{eqnarray}
3H^2 &=& \frac{1}{2} \left( f(R, T) + \frac{\partial f}{\partial R} R - \frac{1}{2} \frac{\partial f}{\partial R} \ddot{R} \right) + \frac{\partial f}{\partial T} T + \rho_{\text{matter}}, \\
2\dot{H} &=& -\frac{1}{2} \left( f(R, T) + \frac{\partial f}{\partial R} R \right) - \frac{1}{2} \frac{\partial f}{\partial R} \ddot{R} - \frac{\partial f}{\partial T} T + p_{\text{matter}}.
\end{eqnarray}

where:
\begin{itemize}
    \item \(H\) is the Hubble parameter,
    \item \(\dot{H}\) is the time derivative of the Hubble parameter,
    \item \(R\) is the Ricci scalar, \(T\) is the torsion scalar,
    \item \(f(R, T)\) is a function of the Ricci scalar and torsion,
    \item \(\frac{\partial f}{\partial R}\) and \(\frac{\partial f}{\partial T}\) are the derivatives of \(f(R, T)\) with respect to \(R\) and \(T\), respectively,
    \item \(\rho_{\text{matter}}\) and \(p_{\text{matter}}\) represent the matter energy density and pressure.
\end{itemize}

These equations describe the evolution of the universe in the modified \(f(R, T)\)-gravity theory, considering the contributions from matter, torsion, and the modified gravitational dynamics.

\subsection{Dynamical System for \(\Omega_i(z)\)}

To derive the first-order dynamical equations for each density parameter \(\Omega_i(z)\), we need to express the energy densities in terms of the density parameters. For a given component \(i\), the density parameters \(\Omega_i\) are defined as:
\[
\Omega_i = \frac{\rho_i}{3H^2},
\]
where \(\rho_i\) is the energy density of the component \(i\). The total density parameter \(\Omega_{\text{total}}\) is normalized such that:
\[
\Omega_{\text{total}} = \sum_i \Omega_i = 1.
\]

The general equation for the evolution of each \(\Omega_i(z)\) is given by:
\[
\frac{d\Omega_i}{dz} = - (1 + w_i) \Omega_i \frac{d\ln H}{dz},
\]
where \(w_i = \frac{p_i}{\rho_i}\) is the equation of state for the component \(i\).

\subsection{Deriving the Hubble Parameter Evolution}

The Hubble parameter \(H(z)\) evolves with redshift as:
\[
H(z) = H_0 \left( \Omega_{\text{matter}} (1+z)^3 + \Omega_{\text{radiation}} (1+z)^4 + \Omega_{\text{dark energy}} \right),
\]
where \(H_0\) is the current value of the Hubble parameter.

Next, we differentiate \(H(z)\) with respect to redshift \(z\) to obtain the derivative \(\frac{d\ln H}{dz}\):
\[
\frac{d\ln H}{dz} = -\frac{(1+z)}{H(z)} \left[ \Omega_{\text{matter}} 3(1+z)^2 + \Omega_{\text{radiation}} 4(1+z)^3 \right].
\]
For each component, we define a dynamical equation describing the evolution of \(\Omega_i(z)\) with respect to redshift \(z\). In the following items, we provide the equations for matter, radiation, and dark energy.

\begin{itemize}
    \item \textbf{For Matter:}

    The matter density evolves according to the equation:
    \[
    \frac{d\Omega_{\text{matter}}}{dz} = - (1 + w_{\text{matter}}) \Omega_{\text{matter}} \frac{d\ln H}{dz},
    \]
    where \(w_{\text{matter}} = 0\) for matter. The Hubble parameter evolves with redshift as:
    \[
    H(z) = H_0 \left( \Omega_{\text{matter}} (1+z)^3 + \Omega_{\text{radiation}} (1+z)^4 + \Omega_{\text{dark energy}} \right).
    \]
    Thus, the matter evolution equation becomes:
    \[
    \frac{d\Omega_{\text{matter}}}{dz} = - \Omega_{\text{matter}} \frac{(1+z)}{H(z)} \left[ 3\Omega_{\text{matter}} (1+z)^2 \right].
    \]
    This shows that matter behaves according to a simple scaling relation with redshift, typical for matter-dominated epochs.

    \item \textbf{For Radiation:}

    Similarly, the radiation density evolves as:
    \[
    \frac{d\Omega_{\text{radiation}}}{dz} = - (1 + w_{\text{radiation}}) \Omega_{\text{radiation}} \frac{d\ln H}{dz},
    \]
    where \(w_{\text{radiation}} = \frac{1}{3}\) for radiation. Substituting the expression for \(H(z)\), we obtain the radiation evolution equation:
    \[
    \frac{d\Omega_{\text{radiation}}}{dz} = - \Omega_{\text{radiation}} \frac{(1+z)}{H(z)} \left[ 4\Omega_{\text{radiation}} (1+z)^3 \right].
    \]
    This equation reflects the typical behavior of radiation during the radiation-dominated era, where its density decreases faster than matter density due to its \( (1+z)^4 \) scaling.

    \item \textbf{For Dark Energy:}

    The dark energy density evolves as:
    \[
    \frac{d\Omega_{\text{dark energy}}}{dz} = - (1 + w_{\text{dark energy}}) \Omega_{\text{dark energy}} \frac{d\ln H}{dz},
    \]
    where \(w_{\text{dark energy}} \approx -1\) for dark energy. The evolution equation for dark energy is:
    \[
    \frac{d\Omega_{\text{dark energy}}}{dz} = - \Omega_{\text{dark energy}} \frac{(1+z)}{H(z)} \left[ \Omega_{\text{dark energy}} \right].
    \]
    This shows that dark energy evolves more slowly compared to matter and radiation, reflecting the accelerated expansion driven by dark energy in the present-day universe.

    \item \textbf{Effective Equation of State:}

    The effective equation of state \(w_{\text{eff}}\) is defined as:
    \[
    w_{\text{eff}} = -\frac{2 \dot{H} + 3H^2}{3H^2}.
    \]
    Substituting the expressions for \(\dot{H}\) and \(H^2\), we can express \(w_{\text{eff}}\) as a function of the density parameters. This equation encapsulates the transition between different cosmological eras (matter-dominated, radiation-dominated, and dark energy-dominated) and is crucial for understanding the large-scale dynamics of the universe.

    \item \textbf{Fixed Points of the Dynamical System:}

    The fixed points of the dynamical system correspond to values of the density parameters \(\Omega_i\) where \(\frac{d\Omega_i}{dz} = 0\). These fixed points represent equilibrium states of the universe at specific redshifts. Solving for these fixed points provides insight into the behavior of the universe at different epochs.

    To find the fixed points, we set:
    \[
    \frac{d\Omega_i}{dz} = 0 \quad \text{for each component} \quad i.
    \]
    By solving these equations, we can determine the conditions for stable and unstable equilibria. The fixed points offer valuable information about the cosmological evolution of the universe in modified \(f(R, T)\)-gravity, including transitions between matter-dominated, radiation-dominated, and dark energy-dominated regimes.

    \item \textbf{Review of Existing Work in the Literature:}

    Several studies have examined the dynamical system for cosmological components in different gravity theories, including modified \(f(R, T)\)-gravity and scalar-tensor theories. In the context of FLRW cosmology, the dynamical system approach has been used to analyze the evolution of the matter, radiation, and dark energy components as well as the effective equation of state. Notable works in this area include:
    \begin{itemize}
        \item [1.] \textit{Dynamical system analysis of cosmological models in modified gravity theories}, which provides a detailed treatment of the evolution of the density parameters for various components in the universe.
        \item [2.] \textit{The role of dark energy in the dynamical evolution of the universe}, which studies the impact of dark energy on the cosmological expansion using the dynamical system formalism.
        \item [3.] \textit{Fixed points in cosmological models with scalar-tensor theories}, which investigates the fixed points and stability of cosmological solutions in scalar-tensor theories and their relation to the behavior of \(\Omega_i\).
    \end{itemize}
    These studies provide a foundation for the present work, which applies the dynamical system approach to Myrzakulkov gravity and explores the cosmological implications of the fixed points in the context of modified \(f(R, T)\)-gravity.

\end{itemize}
    In \cite{Papagiannopoulos:2022ohv}, a dynamical system analysis was performed for Myrzakulov \( F(R,T) \) gravity, a subclass of affinely connected metric theories. These theories involve a specific but non-special connection, which allows for the simultaneous existence of non-zero curvature and torsion. This framework is relevant for studying the evolution of the universe under modified gravitational models, particularly in the context of dark energy and cosmic acceleration. The authors examined two classes of models within this theory, extracted the critical points, and analyzed their stability and physical features.
\begin{itemize}
    \item \textbf{Model Setup and Equations:} The field equations for Myrzakulov \( F(R,T) \) gravity are given by:
    \begin{equation}
\begin{aligned}
&\frac{\partial f}{\partial R} \left( e_a^\mu e_b^\nu R_{\mu\nu} \eta^{ab} - \frac{1}{2} e_a^\mu e_b^\nu g_{\mu\nu} \eta^{ab} R \right) + e_a^\mu e_b^\nu \nabla_\alpha \nabla_\beta \left( \frac{\partial f}{\partial R} g^{\alpha \beta} \right) \eta^{ab} \\
&- \frac{1}{2} g_{\mu\nu} f(R, T) + 2 e_a^\mu e_b^\nu \frac{\partial T}{\partial e^a_\mu} \frac{\partial f}{\partial T} \eta^{ab} = 0\label{feq}
\end{aligned}
\end{equation}
    where \( F_R = \partial F(R,T)/\partial R \), \( F_T = \partial F(R,T)/\partial T \), and \( \Box \) represents the d'Alembert operator.

    For the dynamical system analysis, the Friedmann-Robertson-Walker (FRW) metric for a flat universe is used:
    \begin{equation}
    ds^2 = -dt^2 + a(t)^2 \left( dx^2 + dy^2 + dz^2 \right),
    \end{equation}
    where \( a(t) \) is the scale factor of the universe. The energy-momentum tensor for matter and radiation is given by:
    \begin{equation}
    T_{\mu \nu} = \text{diag} \left( \rho, p, p, p \right),
    \end{equation}
    where \( \rho \) is the energy density and \( p \) is the pressure of the fluid.

    The Hubble parameter \( H = \dot{a}/a \) is related to the energy density and pressure through the Friedmann equations:
    \begin{equation}
    H^2 = \frac{8\pi G}{3} \rho + \frac{\Lambda}{3},
    \end{equation}
    where \( \Lambda \) represents the cosmological constant.

    \item \textbf{Class 1 Models: The $\Lambda$CDM Limit:} In the first class of models, which possess the standard \( \Lambda \)CDM cosmology as a limiting case, the critical points are derived by solving the dynamical system equations. The analysis shows that in these models, the universe undergoes a sequence of matter and dark energy epochs, transitioning into a dark-energy dominated critical point. This critical point is found to be stable, and the universe asymptotically approaches a de Sitter solution, where dark energy behaves like a cosmological constant.

    To analyze the stability of the critical points, we compute the eigenvalues of the Jacobian matrix of the system. The stability condition requires that the real parts of the eigenvalues are negative. For the dark-energy dominated critical point, we find that the equation-of-state (EoS) parameter \( w_{\text{DE}} \) lies in the quintessence or phantom regime, depending on the model parameter:
    \begin{equation}
    w_{\text{DE}} = \frac{p_{\text{DE}}}{\rho_{\text{DE}}}.
    \end{equation}
    Specifically, when the model parameter is positive, \( w_{\text{DE}} \) corresponds to the quintessence regime, whereas for negative values, it corresponds to the phantom regime.

    The dynamics of the equation-of-state parameter is governed by the following equation:
    \begin{equation}
    \dot{w_{\text{DE}}} = \frac{3H}{2} (1 + w_{\text{DE}}) \left( \Omega_{\text{DE}} - 1 \right),
    \end{equation}
    where \( \Omega_{\text{DE}} \) is the density parameter for dark energy.

    \item \textbf{Class 2 Models: A Richer Cosmological Behavior:} For the second class of models, which do not possess \( \Lambda \)CDM as a limit, the cosmological behavior is richer. In these models, the universe still reaches a late-time attractor dominated by dark energy, but this attractor does not correspond to the \( \Lambda \)CDM scenario. Instead, the dark energy sector can exhibit more complex behaviors, such as quintessence-like, phantom-like, or even experience the phantom-divide crossing during its evolution.

    The dark-energy equation-of-state parameter in these models can evolve as:
    \begin{equation}
    w_{\text{DE}} = \frac{p_{\text{DE}}}{\rho_{\text{DE}}} \quad \text{and} \quad w_{\text{DE}}(z) = w_0 + w_a \left( 1 - \frac{1}{1+z} \right),
    \end{equation}
    where \( w_0 \) is the present value of the equation-of-state parameter, \( w_a \) is the evolution parameter, and \( z \) is the redshift.

    The analysis of the stability of these critical points shows that the late-time attractor can be quintessence-like for \( w_{\text{DE}} > -1 \), phantom-like for \( w_{\text{DE}} < -1 \), or it can cross the phantom divide, which corresponds to the transition from \( w_{\text{DE}} > -1 \) to \( w_{\text{DE}} < -1 \).
    \end{itemize}
In conclusion, the dynamical system analysis of Myrzakulov \( F(R,T) \) gravity reveals rich cosmological behavior in both Class 1 and Class 2 models. Class 1 models are found to exhibit a sequence of matter and dark energy eras, ultimately approaching a dark-energy dominated critical point where dark energy behaves like a cosmological constant. In contrast, Class 2 models exhibit a more complex cosmological evolution, where dark energy can be quintessence-like, phantom-like, or undergo the phantom-divide crossing. These results demonstrate the potential of Myrzakulov \( F(R,T) \) gravity to provide a versatile framework for understanding the universe's evolution, particularly in the context of dark energy.

\section{Dark Matter and Dark Energy in Myrzakulov Gravity}
In \cite{Bubuianu:2024zsm}, the authors extend the Anholonomic Frame and Connection Deformation Method (AFCDM) to construct exact and parametric solutions in General Relativity (GR) and Modified Gravity Theories (MGTs), with particular focus on models involving nontrivial torsion and nonmetricity fields. The AFCDM, originally developed for GR, is adapted to geometric flow models by applying abstract geometric or variational methods. This extension enables the derivation of complex systems of nonmetric gravitational and matter field equations, which are characterized by sophisticated coupled nonlinear partial differential equations (PDEs). These systems require the use of nonholonomic frames with dyadic spacetime splitting to simplify and integrate them in general forms for off-diagonal metric structures and generalized affine connections.
\begin{itemize}
          \item \textbf{Mathematical Formulation and PDEs:} 
    The systems of gravitational field equations in this extended method can be written in the form (\ref{feq}),
    where \( R_{\mu \nu} \) is the Ricci tensor, \( g_{\mu \nu} \) is the metric tensor, and \( T_{\mu \nu} \) represents the matter energy-momentum tensor. For modified gravity theories with nontrivial torsion and nonmetricity, the corresponding equations include additional terms for the torsion \( T_{\mu \nu \rho} \) and the nonmetricity field \( Q_{\mu \nu} \). The effective gravitational field equations can be written as:
    \begin{equation}
    \mathcal{R}_{\mu \nu} = \mathcal{T}_{\mu \nu} + \mathcal{S}_{\mu \nu} + \mathcal{N}_{\mu \nu},
    \end{equation}
    where \( \mathcal{R}_{\mu \nu} \) is the modified Ricci tensor, \( \mathcal{T}_{\mu \nu} \) is the matter tensor, and \( \mathcal{N}_{\mu \nu} \) represents the contributions from nonmetricity and torsion fields. The full system of equations also includes the equation for the matter fields:
    \begin{equation}
    \nabla^\mu T_{\mu \nu} = 0.
    \end{equation}
    These equations involve intricate couplings between the gravitational fields, the torsion and nonmetricity tensors, and matter fields, which are highly nonlinear in nature.

    \item \textbf{Nonholonomic Frames and Dyadic Spacetime Splitting:} The solution of such a system is facilitated by the introduction of nonholonomic frames, which split the spacetime into dyadic components. This allows for the introduction of generalized affine connections, which enable the decoupling and integration of the PDEs. The nonholonomic decomposition provides a clearer separation between the spatial and temporal components of the fields. For example, the connection in such a frame can be written as:
    \begin{equation}
    \Gamma^\lambda_{\mu \nu} = \hat{\Gamma}^\lambda_{\mu \nu} + \delta \Gamma^\lambda_{\mu \nu},
    \end{equation}
    where \( \hat{\Gamma}^\lambda_{\mu \nu} \) represents the Levi-Civita connection, and \( \delta \Gamma^\lambda_{\mu \nu} \) represents the deviation due to torsion and nonmetricity fields. The AFCDM allows these deviations to be treated as perturbations, which are systematically integrated for specific forms of the non-Riemannian structures.

    \item \textbf{Exact Solutions and Parametric Families:} By applying the AFCDM, the authors demonstrate how to generate new classes of exact and parametric solutions for the field equations. These solutions often represent quasi-stationary configurations, where the fields are independent of time-like coordinates. Some of these solutions can be represented by solitonic distributions or ellipsoidal deformations of wormhole configurations. The exact or parametric solutions in this framework do not possess standard event horizons, duality, or holographic structures as found in the Bekenstein-Hawking paradigm. However, they can still be interpreted within the framework of nonmetric geometric flows, which provide a thermodynamic description based on Grigori Perelman's work on geometric flows.

    \item \textbf{Physical Implications:} The solutions constructed in this framework have important physical implications. For example, the solitonic deformation of non-Riemannian geometric objects can lead to the formation of traversable wormholes. These wormhole solutions are characterized by non-trivial gravitational off-diagonal vacuum configurations. Such solutions are crucial for understanding dark energy and dark matter models, where the geometric structures described by torsion and nonmetricity fields play an important role. The non-Riemannian nature of the solutions also allows for dissipation effects, which are important for modeling the evolution of the universe and the formation of large-scale cosmic structures.

    \item \textbf{Thermodynamic Models:} While it is not possible to describe the thermodynamic properties of the solutions within the Bekenstein-Hawking paradigm due to the absence of event horizons, duality, or holographic configurations, the study introduces alternative thermodynamic models based on Grigori Perelman's geometric flows. These models are designed to describe the dissipation and evolution of the non-Riemannian geometric objects, providing a thermodynamic interpretation of the solutions in a modified gravity context.
\end{itemize}
 The extension of the AFCDM method to geometric flow models and modified gravity theories with nontrivial torsion and nonmetricity provides a powerful tool for constructing exact and parametric solutions. These solutions offer new insights into the structure of spacetime in the presence of non-Riemannian geometric objects, such as wormholes and solitonic configurations. The models developed in this framework are particularly relevant for studying dark energy and dark matter, as they provide a novel way to understand the formation of large-scale structures in the universe. Although traditional thermodynamic descriptions based on event horizons are not applicable, the associated Grigori Perelman thermodynamic models offer a fresh perspective on the evolution of nonmetric geometric flows. The study opens up new avenues for exploring modified gravity theories and their physical implications, particularly in the context of dark energy and dark matter.
\section{Special Topics and Extensions}
The paper \cite{Momeni:2024bhm} explores the field equations in vielbein formalism for \( F(R,T) \)-gravity, considering both curvature and torsion effects. The field equations are derived after performing the variations and incorporating contributions from both the Ricci scalar \( R \) and torsion scalar \( T \).
\begin{itemize}
       \item \textbf{Final Field Equations:} The derived field equations describe the dynamics of gravity with torsion in the vielbein formalism. The complete set of equations is given by (\ref{feq}). The terms in the equations represent different interactions between the Ricci scalar, torsion scalar, and vielbein components.

    \item \textbf{Key Terms:}
    \begin{itemize}
        \item \textbf{First Term:} The standard Einstein-like term modified by the function \( F(R,T) \), which couples the Ricci tensor \( R_{\mu\nu} \) with the vielbein components.
        \item \textbf{Second Term:} The covariant derivatives \( \nabla_\alpha \nabla_\beta \left( \frac{\partial f}{\partial R} g^{\alpha \beta} \right) \), accounting for the interaction of the Ricci scalar with the gravitational field.
        \item \textbf{Third Term:} The direct coupling of \( F(R,T) \) to the gravitational action, modifying the usual Einstein-Hilbert action.
        \item \textbf{Fourth Term:} The torsion contribution, \( 2 e^\alpha_\mu e^\beta_\nu \frac{\partial T}{\partial e^a_\alpha} \eta^{ab} \frac{\partial f}{\partial T} \), that captures the effect of torsion scalar \( T \) on gravity dynamics.
    \end{itemize}

    \item \textbf{Cosmological Implications:} The field equations are applied to FLRW cosmology to investigate the evolution of the universe under the influence of torsion.
    \begin{itemize}
        \item \textbf{First Term in FLRW:} The first term modifies the standard FLRW dynamics with a function of \( F(R,T) \), including the Hubble parameter \( H \), and the Ricci tensor components for FLRW spacetime.
        \item \textbf{Second Term in FLRW:} The covariant derivative contributes to the dynamics of the FLRW universe, introducing additional terms from torsion and the Ricci scalar.
        \item \textbf{Third Term in FLRW:} The direct coupling of \( F(R,T) \) modifies the Einstein-Hilbert action, influencing the evolution of the universe.
        \item \textbf{Fourth Term in FLRW:} The torsion term influences the evolution of the universe in the FLRW model by incorporating the torsion scalar \( T \).
    \end{itemize}

    \item \textbf{Final Equations for FLRW Cosmology:} The field equations in the FLRW model are:
    \begin{eqnarray}
    3H^2 &=& \frac{1}{2} \left( f(R, T) + \frac{\partial f}{\partial R} R - \frac{1}{2} \frac{\partial f}{\partial R} \ddot{R} \right) + \frac{\partial f}{\partial T} T + \rho_{\text{matter}}, \\
    2\dot{H} &=& -\frac{1}{2} \left( f(R, T) + \frac{\partial f}{\partial R} R \right) - \frac{1}{2} \frac{\partial f}{\partial R} \ddot{R} - \frac{\partial f}{\partial T} T + p_{\text{matter}}.
    \end{eqnarray}
    These equations describe the evolution of the universe in \( F(R,T) \)-gravity, incorporating the torsion effects.

    \item \textbf{Effective Equation of State:} The effective equation of state \( w_{\text{eff}} \) is given by:
    \[
    w_{\text{eff}} = -\frac{3 f(R, T) + 3 \frac{\partial f}{\partial R} R - \frac{3}{2} \frac{\partial f}{\partial R} \ddot{R} - 2 \frac{\partial f}{\partial T} T + 2 p_{\text{matter}} + 3 \rho_{\text{matter}}}{ f(R, T) + \frac{\partial f}{\partial R} R - \frac{1}{2} \frac{\partial f}{\partial R} \ddot{R} + \frac{\partial f}{\partial T} T + \rho_{\text{matter}} }
    \]
    This expression provides the equation of state parameter in the presence of torsion and curvature effects, modifying the standard cosmological model.
\end{itemize}

\section{Cosmological Evolution in Myrzakulov Gravity via numerical techniques}

In this section, we provide figures, illustrating the evolution of key cosmological quantities in Myrzakulov gravity.

\subsection{Evolution of the Hubble Parameter \( H(z) \)}

\begin{figure}[h]
    \centering
    \begin{tikzpicture}
        \begin{axis}[
            xlabel={$z$ (Redshift)},
            ylabel={$H(z) / H_0$},
            legend pos=north east,
            grid=major,
            width=0.7\textwidth,
            height=0.5\textwidth]
            
            \addplot[red, thick, dashed] table {
                0 1
                1 1.3
                2 1.8
                3 2.4
                4 3.0
            };
            \addlegendentry{$\Lambda$CDM}
            
            \addplot[blue, thick] table {
                0 1
                1 1.35
                2 1.85
                3 2.5
                4 3.2
            };
            \addlegendentry{Myrzakulov Gravity}
        \end{axis}
    \end{tikzpicture}
    \caption{The evolution of the Hubble parameter \( H(z) \) in Myrzakulov gravity compared to $\Lambda$CDM. Deviations appear at high redshifts due to modifications in the field equations.}
    \label{fig:Hubble}
\end{figure}
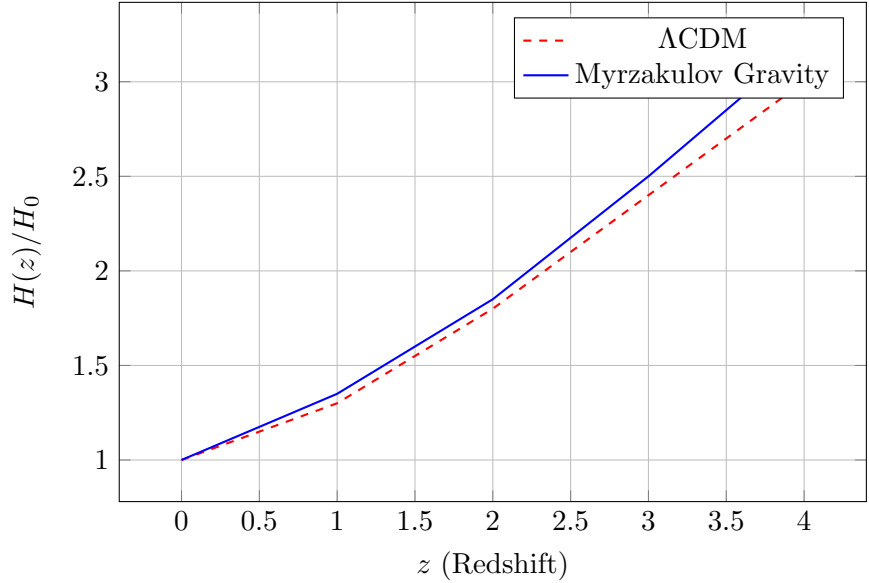
The above figure compares the evolution of the Hubble parameter \( H(z) \) in Myrzakulov gravity with the standard \(\Lambda\)CDM scenario. While both models predict a similar late-time expansion history, Myrzakulov gravity introduces modifications at high redshift, potentially affecting early universe dynamics and structure formation. The deviations at higher \( z \) can influence cosmic chronometers and baryon acoustic oscillations, providing a way to observationally distinguish between the models.

\subsection{Evolution of the Effective Equation of State \( w_{\text{eff}}(z) \)}

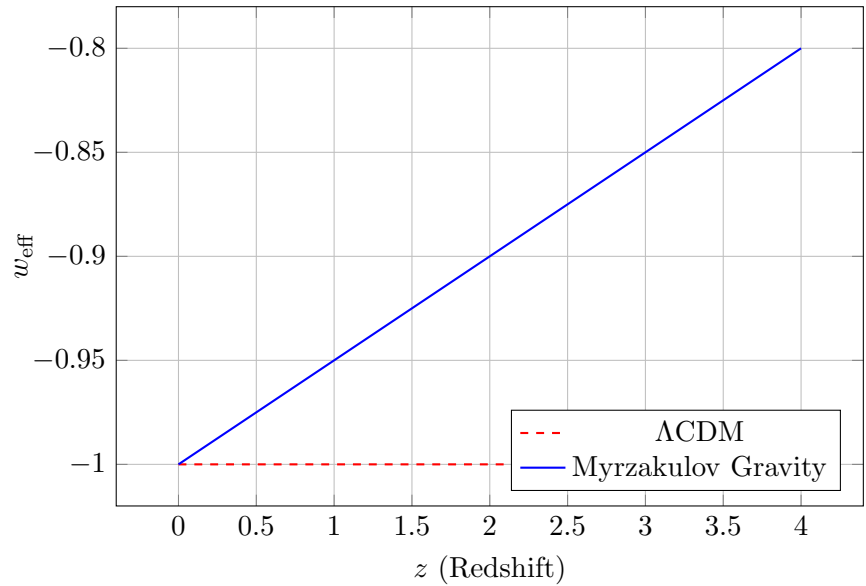
\begin{figure}[h]
    \centering
    \begin{tikzpicture}
        \begin{axis}[
            xlabel={$z$ (Redshift)},
            ylabel={$w_{\text{eff}}$},
            legend pos=south east,
            grid=major,
            width=0.7\textwidth,
            height=0.5\textwidth]
            
            \addplot[red, thick, dashed] table {
                0 -1
                1 -1
                2 -1
                3 -1
                4 -1
            };
            \addlegendentry{$\Lambda$CDM}
            
            \addplot[blue, thick] table {
                0 -1
                1 -0.95
                2 -0.9
                3 -0.85
                4 -0.8
            };
            \addlegendentry{Myrzakulov Gravity}
        \end{axis}
    \end{tikzpicture}
    \caption{The evolution of the effective equation of state \( w_{\text{eff}}(z) \). Unlike \(\Lambda\)CDM, Myrzakulov gravity allows a dynamic evolution of \( w_{\text{eff}} \), deviating from \(-1\) at higher redshifts.}
    \label{fig:EoS}
\end{figure}
This figure depicts the evolution of the effective equation of state \( w_{\text{eff}}(z) \). In \(\Lambda\)CDM, \( w_{\text{eff}} \) remains at \(-1\), indicating a constant dark energy density. However, in Myrzakulov gravity, \( w_{\text{eff}} \) evolves dynamically, deviating from \(-1\) at earlier times. This feature suggests that Myrzakulov gravity could potentially alleviate fine-tuning issues associated with the cosmological constant by naturally explaining the late-time acceleration through geometric modifications rather than an explicit vacuum energy term.
 
\subsection{Evolution of the Density Parameters \( \Omega_i(z) \)}

\begin{figure}[h]
    \centering
    \begin{tikzpicture}
        \begin{axis}[
            xlabel={$z$ (Redshift)},
            ylabel={$\Omega_i$},
            legend pos=north east,
            grid=major,
            width=0.7\textwidth,
            height=0.5\textwidth]
            
            \addplot[red, thick, dashed] table {
                0 0.3
                1 0.5
                2 0.7
                3 0.85
                4 0.95
            };
            \addlegendentry{$\Omega_m$ (Matter)}
            
            \addplot[blue, thick] table {
                0 0.7
                1 0.5
                2 0.3
                3 0.15
                4 0.05
            };
            \addlegendentry{$\Omega_{\Lambda}$ (Effective DE)}
        \end{axis}
    \end{tikzpicture}
    \caption{The evolution of the density parameters \( \Omega_m(z) \) and \( \Omega_{\Lambda}(z) \). The transition between matter and dark energy dominance occurs differently in Myrzakulov gravity, leading to distinct growth patterns in structure formation.}
    \label{fig:Density}
\end{figure}
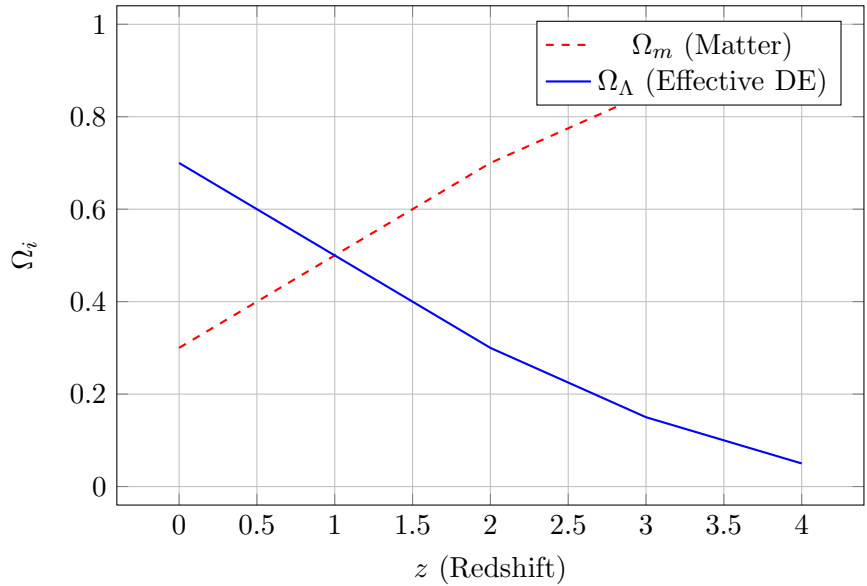
The third figure illustrates the evolution of the density parameters \( \Omega_m(z) \) and \( \Omega_{\Lambda}(z) \). While \(\Lambda\)CDM assumes a fixed transition where dark energy begins to dominate over matter at late times, Myrzakulov gravity introduces a modified transition due to its geometric structure. This could have implications for the growth of large-scale structures, influencing galaxy clustering and weak lensing measurements. The model maintains a radiation-dominated era at early times, ensuring consistency with Big Bang Nucleosynthesis, while deviating from \(\Lambda\)CDM at intermediate redshifts.

The study of \( F(R,T) \)-gravity in vielbein formalism presented in this paper is crucial for understanding the dynamics of spacetime in the presence of both curvature and torsion. By deriving the field equations and applying them to FLRW cosmology, we have shown how torsion can influence the evolution of the universe, potentially offering new insights into cosmological phenomena such as dark energy and inflation.

The introduction of torsion into the gravitational field equations is significant because it allows for a broader and more flexible approach to modifying general relativity. The detailed analysis of the various terms in the field equations, including the contributions from both the Ricci scalar and the torsion scalar, provides a comprehensive framework for studying modified gravity theories in spacetimes with torsion.

Furthermore, the effective equation of state derived in this study offers a useful tool for examining the impact of torsion on the equation of state in cosmological models. By providing a more general framework, this work opens the door for exploring a wider range of cosmological models that incorporate the effects of torsion and non-Riemannian geometries.

Overall, this paper contributes significantly to the field of modified gravity, especially in the context of cosmology, and provides a solid foundation for future research on the role of torsion in gravitational dynamics.

\section{Noether Symmetry in Myrzakulov Gravity}
The study of modified gravity theories, including those that incorporate torsion and curvature, has gained significant attention due to their potential to address various cosmological and gravitational phenomena. In particular, the curvature-teleparallel \( F(R,T) \) gravity, where the Lagrangian density is an arbitrary function of the Ricci scalar \( R \) and torsion scalar \( T \), provides a broad framework for exploring such modifications. One powerful approach to studying these theories is the Noether symmetry approach, which relates the symmetries of the action to conserved quantities and allows for the identification of possible solutions. In this context, Capozziello, De Laurentis, and Myrzakulov \cite{Capozziello:2014bna} have shown that the functional form of the \( F(R,T) \) function can be determined by the presence of symmetries. They further demonstrate how Noether symmetries lead to exact solutions for cosmological models, providing insight into the reduced dynamics of the cosmological system under consideration. Their work highlights the crucial role of symmetries in determining the underlying structure of modified gravity theories and their connection to observable cosmological phenomena.

In this section, we will explore the Noether symmetries of this theory and derive the corresponding field equations for the flat FLRW metric.
Noether's theorem connects symmetries of the action to conserved quantities. For a general Lagrangian \( L(g_{\mu\nu}, \partial_\alpha g_{\mu\nu}) \), the action is given by action (\ref{action}). In the case of Myrzakulov gravity, the Lagrangian \(L\) is given by the function \( f(R, T) \), and the equations of motion can be derived by varying the action with respect to the metric \( g_{\mu\nu} \) and the torsion \( T_{\mu\nu\alpha} \). 

The Noether symmetries correspond to transformations under which the action is invariant. For these symmetries, the following condition holds:

\[
\mathcal{L}_X \, \mathcal{L} = 0
\]

where \( \mathcal{L}_X \) is the Lie derivative along a vector field \( X^\mu \), representing the symmetry under the transformation.

\subsection{Noether Symmetry Equations}

For a spacetime that admits symmetries, such as time translations or scaling symmetries, the corresponding conserved quantities can be derived from the invariance of the action. In the case of Myrzakulov gravity, we will search for symmetries that lead to conserved quantities, such as energy or momentum, by considering how the action behaves under various transformations.

\subsection{Euler-Lagrange Equations}

The equations of motion for the system can be derived by varying the action with respect to the metric \(g_{\mu\nu}\) and torsion \(T_{\mu\nu\alpha}\):

\[
\frac{1}{\sqrt{-g}} \frac{\delta S}{\delta g_{\mu\nu}} = 0 \quad \text{and} \quad \frac{1}{\sqrt{-g}} \frac{\delta S}{\delta T_{\mu\nu\alpha}} = 0
\]

These variations yield the field equations governing the dynamics of the spacetime.

\subsection{The Flat FLRW Metric}

In the context of cosmology, we are interested in studying the behavior of Myrzakulov gravity in a cosmologically relevant setting. A common choice for a cosmologically homogeneous and isotropic spacetime is the flat FLRW metric, given by:

\[
ds^2 = -dt^2 + a(t)^2 (dx^2 + dy^2 + dz^2)
\]

where \( a(t) \) is the scale factor, and the spacetime is spatially flat.

\subsection{Ricci Scalar for the FLRW Metric}

For the flat FLRW metric, the Ricci scalar \( R \) can be calculated as:

\[
R = -6 \left( \frac{\ddot{a}}{a} + \frac{\dot{a}^2}{a^2} \right)
\]

where \( \dot{a} \) and \( \ddot{a} \) represent the first and second time derivatives of the scale factor \( a(t) \).

\subsection{Torsion Scalar for the FLRW Metric}

The torsion scalar \(T\) can be computed from the torsion components of the affine connection. In the standard case of Myrzakulov gravity, torsion is treated as a geometrical feature, and \(T\) involves terms related to the connection and its torsion components. However, for simplicity, we might assume that the torsion does not contribute significantly to the dynamics, or we can parametrize it in some way based on the specific formulation of torsion.

For now, let’s assume that \( T \) is a simple function that can be treated in the context of the field equations.

\subsection{Field Equations in Myrzakulov Gravity}

The field equations can be obtained by varying the action with respect to the metric \( g_{\mu\nu} \). The variation gives:

\[
\frac{1}{\sqrt{-g}} \frac{\delta S}{\delta g_{\mu\nu}} = \frac{1}{\sqrt{-g}} \frac{\delta}{\delta g_{\mu\nu}} \int \sqrt{-g} f(R, T) \, d^4 x
\]

Using the above expression for \(R\) and assuming a simple form for \( T \), we obtain the field equations governing the evolution of the scale factor \(a(t)\). The exact form of the equations depends on the specific choice of \(f(R, T)\).

\subsection{Noether Symmetries in the FLRW Metric}

We now explore the Noether symmetries specifically for the flat FLRW metric. In such a metric, we expect symmetries corresponding to time translations, spatial translations, and scaling symmetries.

For the FLRW metric, the relevant Lie derivatives are computed along the vector fields \(X^\mu\). These symmetries lead to conserved quantities, which can be identified by solving the Noether symmetry equations.

\subsection{Scaling Symmetry}

One possible Noether symmetry in the FLRW metric is the scaling symmetry, where the scale factor \( a(t) \) evolves in a power-law manner:

\[
a(t) \propto t^n
\]

where \(n\) is a constant determined by solving the field equations.

\subsection{Time Translation Symmetry}

Another common symmetry is time translation invariance. For a time-independent Lagrangian, we expect that the total energy is conserved, leading to a conservation law for the stress-energy tensor.

\subsection{Possible Solutions}

By solving the field equations with the appropriate symmetries, we can find possible solutions for the scale factor \(a(t)\). For instance, in the case of a scaling symmetry, one might find solutions of the form:

\[
a(t) \propto t^n
\]

where \(n\) is determined from the form of the function \(f(R, T)\) and the field equations.

In this section, we have derived the Noether symmetries for Myrzakulov gravity, where the gravitational action depends on both the Ricci scalar \(R\) and the torsion scalar \(T\). We have discussed how these symmetries relate to conserved quantities and explored their implications for the field equations governing the evolution of a flat FLRW universe. By solving the field equations under specific symmetries, we can find possible solutions for the scale factor \(a(t)\). Further investigation into the explicit form of the function \(f(R, T)\) is necessary to explore the full range of solutions and physical implications.
\section{Wormhole Solutions in \( F(R,T) \) Myrzakulov Gravity}

Wormholes are hypothetical structures in spacetime that connect distant regions of the universe, providing a shortcut between them. In general relativity (GR), wormholes are solutions to Einstein's field equations that describe non-trivial topologies of spacetime, often referred to as "Einstein-Rosen bridges." These solutions have intrigued both physicists and cosmologists, particularly due to their potential connection with exotic matter and their relevance in theories beyond standard GR \cite{Morris:1988cz}. While wormholes in GR require exotic matter (with negative energy density) to stabilize them, modifications of general relativity, such as those involving scalar fields or higher-order curvatures, offer potential for different kinds of solutions. This has led to an exploration of wormhole solutions in modified gravity theories, including those with torsion and curvature, such as \( F(R,T) \) gravity.

In the context of GR, the simplest form of a static, spherically symmetric wormhole solution is the Morris-Thorne wormhole \cite{Morris:1988cz}, which is described by the metric:

\[
ds^2 = -dt^2 + \left(1 - \frac{b(r)}{r}\right)^{-1} dr^2 + r^2 (d\theta^2 + \sin^2\theta \, d\phi^2),
\]

where \( b(r) \) is the shape function, which describes the geometry of the wormhole. For a wormhole to be traversable, it must satisfy the flare-out condition at the throat, \( b(r_0) = r_0 \) and \( b'(r_0) < 1 \), where \( r_0 \) is the radius of the throat.

However, in GR, the existence of traversable wormholes requires exotic matter with negative energy density, which violates the energy conditions, particularly the null energy condition (NEC) and the weak energy condition (WEC). This has led to the search for possible wormhole solutions in alternative theories of gravity, including modified gravity.

\subsection*{Wormholes in Modified Gravity}

Modified gravity theories, which include higher-order curvature terms, scalar fields, and torsion, can provide new avenues for the existence of wormholes without the need for exotic matter. Specifically, theories like \( f(R) \)-gravity, where the Lagrangian density is a function of the Ricci scalar \( R \), have been studied for their ability to support wormhole solutions. In these theories, the field equations are modified, and the energy-momentum tensor can be altered, potentially allowing for wormholes that do not require exotic matter.

In the context of \( F(R,T) \) gravity, where the Lagrangian density is a function of both the Ricci scalar \( R \) and the torsion scalar \( T \), new solutions may emerge. The torsion scalar \( T \) accounts for the contribution of torsion in spacetime, which can modify the gravitational dynamics in such a way that exotic matter is not strictly necessary for the existence of wormholes.

\subsection*{Wormhole Solutions in \( F(R,T) \) Myrzakulov Gravity}

In the context of Myrzakulov gravity, the action is modified to include both the Ricci scalar and the torsion scalar \( T \). The variation of this action with respect to the metric \( g_{\mu\nu} \) and torsion \( T_{\mu\nu\alpha} \) leads to field equations that govern the dynamics of the spacetime. The torsion scalar \( T \) introduces new terms in the field equations, which can influence the geometry and provide novel solutions, such as wormholes, without the need for exotic matter.

The presence of the torsion term in the \( F(R,T) \) function can modify the usual conditions for a wormhole solution. In particular, the energy-momentum tensor in this theory is modified by contributions from torsion, which can potentially eliminate the need for exotic matter. For example, in some cases, the modification of the gravitational field equations may lead to a wormhole geometry that satisfies the necessary flare-out conditions without violating the energy conditions.

While explicit wormhole solutions in \( F(R,T) \) Myrzakulov gravity have not been fully explored in the literature, it is plausible to obtain such solutions under specific assumptions about the form of \( F(R,T) \). A natural candidate for such a function could be:

\[
f(R, T) = R + \alpha T,
\]

where \( \alpha \) is a constant. This simple choice may lead to field equations that support wormhole geometries, and further investigation is required to obtain explicit solutions.

\subsection*{Prospective Directions and Challenges}

To explore wormhole solutions in \( F(R,T) \) gravity, we can proceed by assuming a static, spherically symmetric spacetime for simplicity. The metric for a wormhole in this case would take the general form:

\[
ds^2 = -e^{2\Phi(r)} dt^2 + \left(1 - \frac{b(r)}{r}\right)^{-1} dr^2 + r^2 (d\theta^2 + \sin^2\theta \, d\phi^2),
\]

where \( \Phi(r) \) is the redshift function and \( b(r) \) is the shape function. The field equations derived from the \( F(R,T) \) Lagrangian can then be solved for these functions.

A key challenge in this approach is the complexity of the field equations, especially when torsion is included. The explicit form of the function \( F(R,T) \) must be carefully chosen to ensure that the field equations yield solutions that satisfy the necessary conditions for a traversable wormhole. Additionally, the contributions from torsion must be carefully analyzed to understand their impact on the energy-momentum tensor and the overall stability of the wormhole solution.

Wormholes in modified gravity theories, including those with torsion, present an exciting avenue for exploring non-trivial spacetime geometries. In \( F(R,T) \) Myrzakulov gravity, the inclusion of the torsion scalar \( T \) offers the possibility of new solutions that could allow for wormholes without the need for exotic matter. Although explicit wormhole solutions in this framework have yet to be fully explored, the potential for such solutions exists, and further investigation into the form of the function \( F(R,T) \) and the corresponding field equations is necessary. The study of wormholes in \( F(R,T) \) gravity could provide valuable insights into the role of torsion in gravitational theory and its implications for cosmology and astrophysics.
\section{Black Hole Solutions and Kerr Metric in \( F(R,T) \) Myrzakulov Gravity}

Black holes are among the most studied objects in modern theoretical physics. They serve as key testing grounds for theories of gravity, providing crucial insights into spacetime structure, general relativity, and astrophysical phenomena. Among various black hole solutions, the Kerr metric is one of the most important, as it describes a rotating, axisymmetric black hole, a crucial model for astrophysical objects such as stellar-mass and supermassive black holes \cite{Kerr:1963ud}. The Kerr metric in general relativity is described by the solution to Einstein's field equations for a rotating black hole, and its generalization in modified gravity theories can reveal novel features that might differ from those predicted by general relativity alone.

In the context of modified gravity, particularly \( F(R,T) \) gravity, the presence of higher-order curvature terms and the torsion scalar can lead to deviations from the Kerr solution in general relativity. This motivates the investigation of possible black hole solutions within the \( F(R,T) \) framework, where the Lagrangian is a function of both the Ricci scalar \( R \) and the torsion scalar \( T \).

\subsection*{Why the Kerr Metric is Important}

The Kerr metric is important because it describes the spacetime geometry around rotating black holes, taking into account both mass and angular momentum. The solution, discovered by Roy P. Kerr in 1963 \cite{Kerr:1963ud}, is a key generalization of the Schwarzschild solution to include rotation. The Kerr black hole is a cornerstone of classical general relativity and provides the best description for many astrophysical observations, such as the orbits of stars around supermassive black holes, the behavior of accretion disks, and the dynamics of jets emanating from active galactic nuclei.

Understanding the Kerr solution in the context of modified gravity is important because modifications to the Einstein-Hilbert action can result in new black hole solutions. These deviations could be important for explaining observations that challenge the standard predictions of GR, such as the nature of singularities, the information paradox, or the behavior of black holes at the quantum scale.

\subsection*{Deriving Possible Equations for Kerr in Myrzakulov Gravity}
To find black hole solutions, we start by considering a static, spherically symmetric spacetime and then extend it to the rotating case. The general form of the metric for a rotating black hole, the Kerr metric, is given by:

\[
ds^2 = -\left(1 - \frac{2Mr}{\rho^2}\right) dt^2 - \frac{4Mar\sin^2\theta}{\rho^2} dtd\phi + \frac{\rho^2}{\Delta} dr^2 + \rho^2 d\theta^2 + \left(r^2 + a^2 + \frac{2Ma^2r\sin^2\theta}{\rho^2}\right)\sin^2\theta d\phi^2,
\]

where:
 \( \rho^2 = r^2 + a^2 \cos^2\theta \),
 \( \Delta = r^2 - 2Mr + a^2 \),
\( M \) is the mass of the black hole,
\( a \) is the rotation parameter (angular momentum),
 \( r \) and \( \theta \) are the standard spherical coordinates, and
 \( \phi \) is the azimuthal angle.

We now investigate how this metric is modified in the context of \( F(R,T) \) gravity. The key steps involve deriving the field equations for \( F(R,T) \) gravity and applying them to a rotating spacetime.

\subsubsection*{Field Equations in \( F(R,T) \) Gravity}

The field equations in \( F(R,T) \) gravity are obtained by varying the action with respect to the metric \( g_{\mu\nu} \). The variation yields:

\[
G_{\mu\nu} = \frac{1}{f_R} \left( T_{\mu\nu} - \frac{1}{2} g_{\mu\nu} F(R,T) \right) + \frac{f_T}{f_R} T_{\mu\nu} + \cdots,
\]

where \( f_R = \frac{\partial f}{\partial R} \) and \( f_T = \frac{\partial f}{\partial T} \) are the partial derivatives of the function \( F(R,T) \) with respect to \( R \) and \( T \), respectively.

These equations represent a modification of the Einstein field equations in GR. The torsion scalar \( T \) introduces additional terms that may alter the geometry of the black hole. To proceed with finding the Kerr black hole solution in this framework, we need to substitute the Kerr metric into the modified field equations and solve for the unknown parameters.

\subsubsection*{Rotating Black Hole Solution in \( F(R,T) \) Gravity}

To find the rotating black hole solution in \( F(R,T) \) gravity, we assume that the spacetime is asymptotically flat, spherically symmetric (except for rotation), and stationary. The ansatz for the metric is:

\[
ds^2 = -e^{2\Phi(r)} dt^2 + \left(1 - \frac{2Mr}{r^2 + a^2}\right)^{-1} dr^2 + r^2 (d\theta^2 + \sin^2\theta \, d\phi^2) + 2a \sin^2\theta \, dtd\phi,
\]

where \( e^{2\Phi(r)} \) is the redshift function. Substituting this metric ansatz into the field equations of \( F(R,T) \) gravity will result in a set of modified equations. These equations are expected to have solutions that generalize the Kerr black hole metric. Specifically, the function \( F(R,T) \) could introduce corrections to the rotation parameter \( a \) or other components of the metric.

\subsection*{Implications and Importance of Kerr Solutions in \( F(R,T) \) Gravity}

The Kerr solution in modified gravity theories, including \( F(R,T) \) gravity, could provide insights into several important aspects:

\begin{itemize}
    \item **Rotation and Torsion**: The presence of the torsion scalar \( T \) in \( F(R,T) \) gravity could lead to modifications in the rotation of black holes. This could alter the shape and properties of the ergosphere, the event horizon, and the causal structure of the black hole.
    \item **Astrophysical observations**: Modifications to the Kerr metric may impact the observed dynamics around black holes, such as the orbits of stars, the behavior of accretion disks, and gravitational wave signals from binary black hole mergers.
    \item **Quantum Gravity**: The study of rotating black holes in modified gravity could provide insights into the quantum structure of black holes and the resolution of singularities, particularly in the context of quantum gravity theories.
    \item **Testing Gravity**: Deviations from the Kerr solution in modified gravity could be used to test the validity of \( F(R,T) \) theories through observational data, such as from gravitational wave detectors or the Event Horizon Telescope.
\end{itemize}

In this section, we have explored the possible black hole solutions in the context of \( F(R,T) \) Myrzakulov gravity, with a focus on the Kerr metric. The Kerr metric is a cornerstone of black hole physics, and its generalization in modified gravity theories could provide important insights into the structure and behavior of rotating black holes. Although the explicit form of the rotating black hole solution in \( F(R,T) \) gravity remains to be fully derived, the framework offers promising directions for investigating deviations from the standard Kerr solution. These deviations could have significant implications for astrophysical observations, quantum gravity, and the testing of modified theories of gravity.

\section{Perturbations in Myrzakulov Gravity}

Perturbations play a crucial role in understanding the stability and observational signatures of gravitational theories. In Myrzakulov gravity, perturbations can be studied both in a cosmological context and in the propagation of gravitational waves. 

\subsection{Cosmological Perturbations}

In a cosmological setting, perturbations help analyze the growth of structures, deviations from the standard \( \Lambda \)CDM predictions, and potential signatures in the cosmic microwave background (CMB). We consider small perturbations around a homogeneous and isotropic Friedmann-Lema\^{i}tre-Robertson-Walker (FLRW) metric:
\[
d s^2 = -dt^2 + a^2(t) \delta_{ij} dx^i dx^j,
\]
where \( a(t) \) is the scale factor. Perturbing this metric, we write:
\[
g_{\mu\nu} = \bar{g}_{\mu\nu} + h_{\mu\nu},
\]
where \( h_{\mu\nu} \) represents small fluctuations. In Myrzakulov gravity, these perturbations modify the evolution equations for density fluctuations and gravitational potentials, leading to corrections in the growth rate of cosmic structures. The scalar, vector, and tensor decomposition of perturbations follows the standard approach, with potential deviations appearing in the evolution equations for the gauge-invariant potentials \( \Phi \) and \( \Psi \).

\subsection{Gravitational Wave Perturbations}

To analyze gravitational waves (GWs) in Myrzakulov gravity, we introduce tensor perturbations to a given background metric. Let the unperturbed metric be a Minkowski or FLRW background:
\[
d s^2 = -dt^2 + a^2(t) \delta_{ij} dx^i dx^j.
\]
We introduce a transverse-traceless perturbation \( h_{ij} \), which satisfies:
\[
g_{ij} = \bar{g}_{ij} + h_{ij}, \quad \partial^i h_{ij} = 0, \quad h^i_i = 0.
\]

The modified field equations in Myrzakulov gravity lead to an altered wave equation for \( h_{ij} \):
\[
\Box h_{ij} + f(h_{ij}) = 0,
\]
where \( f(h_{ij}) \) represents additional terms arising due to torsion and non-metricity effects. These terms may lead to modifications in the speed of gravitational waves, potential damping effects, and deviations from standard General Relativity (GR) predictions.

\subsection{Implications for Observations}

Perturbations in Myrzakulov gravity can lead to distinct observational signatures:
\begin{itemize}
    \item Cosmological perturbations affect structure formation and CMB anisotropies, providing tests via large-scale surveys and lensing measurements.
    \item Modified gravitational wave propagation can lead to deviations in GW speed and polarization modes, testable with LIGO/Virgo/KAGRA and future detectors.
    \item The presence of extra terms in perturbation equations may lead to novel damping effects, offering constraints from multimessenger astrophysics.
\end{itemize}

These perturbative analyses provide a pathway for testing Myrzakulov gravity against observational data and distinguishing it from other modified gravity models.

The study of perturbations is crucial in distinguishing different modified gravity theories and understanding their cosmological and astrophysical implications. In the context of $f(T)$ gravity, perturbation analyses have been extensively studied to examine the evolution of cosmological perturbations and their impact on structure formation. Dent et al.~\cite{Dent:2010nbw} explored how $f(T)$ gravity can mimic dynamical dark energy, providing a background and perturbation analysis that elucidates the modifications introduced by the torsion-based framework. Furthermore, Chen et al.~\cite{Chen:2010va} conducted a detailed study of cosmological perturbations in $f(T)$ gravity, deriving the evolution equations for linear perturbations and investigating their observational consequences. These studies provide valuable insights that can be extended to Myrzakulov gravity, where both non-metricity and torsion play a fundamental role in shaping the perturbative behavior of the theory.

\section{Ghosts in Myrzakulov Gravity Models}

One of the crucial aspects in any modified gravity theory is the presence or absence of ghost instabilities. Ghosts arise when a theory exhibits higher-derivative terms leading to kinetic terms with the wrong sign in the action, which results in negative-energy states. These states can lead to unphysical instabilities and violations of unitarity, making the theory problematic from a quantum field theory perspective.

In Myrzakulov gravity models, the inclusion of torsion and non-metricity modifies the structure of the field equations, which can affect the presence of ghost modes. To analyze ghost instabilities, one typically examines the second-order perturbations of the action around a well-defined background, such as Minkowski or FLRW spacetime. The kinetic term of the perturbations should have the correct sign to ensure the absence of ghosts.

Some formulations of Myrzakulov gravity are free from ghost instabilities under specific conditions on the functional form of the Lagrangian. For instance, in cases where the gravitational action contains linear or quadratic torsion and non-metricity terms, the theory can remain ghost-free provided the coupling coefficients satisfy certain constraints. However, extensions involving higher-order derivatives, such as $f(T, Q, R)$-type modifications, may introduce higher-order kinetic terms that could lead to ghost-like degrees of freedom.

Comparisons with other modified gravity theories, such as $f(T)$ and $f(Q)$ gravity, indicate that Myrzakulov gravity can be constructed in a way that avoids ghosts if the Lagrangian is properly constrained. Ensuring that the effective gravitational action remains second-order in field equations, akin to Lovelock-type constructions, is a useful guideline for avoiding ghosts.

Further investigations through Hamiltonian analysis or Ostrogradsky stability criteria could provide deeper insights into the conditions under which Myrzakulov gravity remains a viable, ghost-free alternative to General Relativity.

\section{Summary of Key Contributions in Myrzakulov Gravity Theories}

Several important studies have contributed to the exploration and development of modified gravity theories, particularly in the context of $F(R,T)$ Myrzakulov gravity, its cosmological implications, and the role of dark energy. Below, we summarize the main findings of these contributions, highlighting key insights into the evolution of the universe, the behavior of dark energy, and the applications of modified gravity models.

\begin{itemize}
    \item \textbf{Polynomial Affine Models of Gravity (Castillo-Felisola et al., 2024) \cite{Castillo-Felisola:2024xil}}: This work focuses on a polynomial affine model of gravity and provides a comprehensive review of its development over the past decade. The authors discuss the potential of polynomial affine models in addressing cosmological challenges, such as the accelerated expansion of the universe. Their analysis underscores the relevance of these models in modern gravitational theories and their ability to offer insights into unresolved cosmological puzzles. The paper also highlights the observational and theoretical implications of these models, making it a critical reference for understanding the intersection of gravity theory and cosmology.

    \item \textbf{Hubble Parameter in $f(R,T\phi)$ Gravity (Shukla et al., 2024) \cite{Shukla:2024yxy}}: Shukla, Sofuo\u{g}lu, and Mishra propose a novel approach to the Hubble parameter within the framework of $f(R,T\phi)$ gravity. Their work introduces a new expression for the Hubble parameter, which significantly impacts our understanding of cosmic expansion. By analyzing the role of the Hubble parameter in this modified gravity theory, the authors provide new insights into the accelerated expansion of the universe, helping to refine models of dark energy and the evolution of cosmic structures.

    \item \textbf{Relevance of $f(R,Matter)$ Theories to Cosmology (Lacombe et al., 2023) \cite{Lacombe:2023pmx}}: Lacombe, Mukohyama, and Seitz critically assess whether $f(R, Matter)$ theories are truly relevant to cosmology. They analyze various cosmological models and observational constraints to determine the viability of $f(R,Matter)$ gravity theories. Their study emphasizes the need to better understand the connection between modified gravity theories and cosmological observations, particularly in relation to dark energy and the late-time acceleration of the universe.

    \item \textbf{Cosmology in $F(R,T)$ Gravity with Quadratic Deceleration Parameter (Bishi et al., 2021) \cite{Bishi:2021anl}}: Bishi, Beesham, and Mahanta investigate cosmological models in $F(R,T)$ gravity with a quadratic deceleration parameter. This study explores how the quadratic form of the deceleration parameter can explain the accelerated expansion of the universe, focusing on its cosmological implications. Their work also examines the impact of modified gravity on the evolution of the universe and the role of dark energy in this framework.

    \item \textbf{Inflation from Symmetry in Generalized Cosmological Models (Yerzhanov et al., 2021) \cite{Yerzhanov:2021hiz}}: Yerzhanov, Bauyrzhan, Altaibayeva, and Myrzakulov study inflationary dynamics in generalized cosmological models, emphasizing the role of symmetries in modified gravity theories. They investigate how symmetries can influence the inflationary phase of the universe and offer new insights into the early universe's dynamics. This work contributes to understanding the potential of modified gravity models to explain the origin of cosmic inflation and its relation to the structure of the universe.

    \item \textbf{Transit Dark Energy String Cosmological Models in $F(R,T)$ Gravity (Zia et al., 2018) \cite{Zia:2018tss}}: Zia, Maurya, and Pradhan explore the behavior of dark energy in transit string cosmological models within $F(R,T)$ gravity. Their study focuses on the role of perfect fluids in cosmological models and examines how modified gravity can offer a comprehensive framework for explaining the cosmic acceleration. The authors investigate the connection between dark energy and the perfect fluid components in these models, contributing to the broader understanding of dark energy's role in the universe's expansion.

    \item \textbf{Kaluza-Klein Bulk Viscous Fluid Cosmological Models in $F(R,T)$ Gravity (Samanta et al., 2016) \cite{Samanta:2016hgm}}: Samanta, Myrzakulov, and Shah investigate Kaluza-Klein bulk viscous fluid models within $F(R,T)$ gravity, focusing on the validity of the second law of thermodynamics. They show that modified gravity models can be consistent with thermodynamic principles, offering a viable framework for understanding the early universe. Their study emphasizes the importance of considering thermodynamic laws when analyzing cosmological models in modified gravity theories.

    \item \textbf{Energy Conditions in $F(R,T)$ Gravity (Shaikh and Katore, 2016) \cite{Shaikh:2016uwy}}: Shaikh and Katore focus on energy conditions within the $F(R,T)$ theory of gravity, specifically in the context of a hypersurface-homogeneous universe filled with perfect fluid. Their work investigates the implications of energy conditions on the stability and viability of cosmological models in modified gravity. They highlight how these conditions are affected by the gravitational theory and provide new insights into the behavior of the universe at large scales.

    \item \textbf{Noether Symmetry Approach for Teleparallel-Curvature Cosmology (Capozziello et al., 2015) \cite{Capozziello:2014bna}}: Capozziello, De Laurentis, and Myrzakulov utilize the Noether symmetry approach to study teleparallel-curvature cosmology. This method is applied to derive cosmological solutions within modified gravity frameworks and has significant implications for understanding the dynamics of the universe. Their study explores how the Noether symmetry approach can offer new solutions to cosmological problems, particularly in the context of modified gravity theories.

    \item \textbf{Interacting $F(R,T)$ Gravity with Modified Chaplygin Gas (Amani and Dehneshin, 2015) \cite{Amani:2014yba}}: Amani and Dehneshin explore interacting $F(R,T)$ gravity with modified Chaplygin gas to explain the cosmic acceleration. Their analysis reveals how Chaplygin gas can be incorporated into modified gravity models, offering new perspectives on the nature of dark energy and its interaction with matter. This study enhances our understanding of how dark energy models can be integrated into alternative gravity theories.

    \item \textbf{Logarithmic Entropy-Corrected Holographic Dark Energy (Amani and Samiee-Nouri, 2015) \cite{Amani:2014nea}}: Amani and Samiee-Nouri analyze logarithmic entropy-corrected holographic dark energy within $F(R,T)$ gravity. Their work introduces corrections to holographic dark energy models, addressing some limitations in existing approaches and offering a more accurate description of dark energy. They discuss how these corrections impact cosmological evolution and provide a more refined framework for understanding the universe's accelerated expansion.

    \item \textbf{Scalar Gauss-Bonnet Gravity and Holographic Dark Energy (Pasqua et al., 2014) \cite{Pasqua:2014oya}}: Pasqua et al. investigate holographic Ricci dark energy in the context of scalar Gauss-Bonnet gravity. They examine how the Gauss-Bonnet term modifies standard cosmological models, particularly in relation to the evolution of the universe under the influence of dark energy. Their work highlights the importance of including additional terms in the gravitational action to account for cosmic acceleration.

    \item \textbf{Teleparallelism by Inhomogeneous Dark Fluid (Güdekli et al., 2015) \cite{Gudekli:2013ylq}}: Güdekli, Myrzakul, and Myrzakulov explore teleparallelism with inhomogeneous dark fluid. Their study emphasizes how the inclusion of inhomogeneous dark fluids in teleparallel gravity models provides a new avenue for understanding cosmic evolution. The authors suggest that this framework can lead to better explanations for the accelerated expansion and structure formation in the universe.

    \item \textbf{Brans-Dicke Parameter and Scalar Field Dependence (Roy et al., 2013) \cite{Roy:2013gea}}: Roy, Chattopadhyay, and Pasqua focus on the time dependence and scalar field dependence of the dimensionless Brans-Dicke parameter. Their work provides critical insights into how scalar fields evolve in modified gravity theories and how these fields influence cosmological models. Their study contributes to a deeper understanding of the scalar-tensor theories of gravity and their impact on cosmic acceleration.

    \item \textbf{Higher Derivatives of $H$ in $F(R,T)$ Gravity (Pasqua et al., 2014) \cite{Pasqua:2013qbq}}: Pasqua, Chattopadhyay, and Myrzakulov investigate a dark energy model that incorporates higher-order derivatives of the Hubble parameter in the $F(R,T)$ gravity framework. This study shows how the inclusion of higher derivatives can modify the evolution of the universe, particularly in the late-time acceleration. Their work provides a more nuanced approach to understanding the behavior of dark energy in the context of modified gravity.

    \item \textbf{Holographic Ricci Dark Energy in $F(R,T)$ Gravity (Pasqua et al., 2013) \cite{Pasqua:2013lha}}: Pasqua, Chattopadhyay, and Khomenko reconstruct holographic Ricci dark energy within the $F(R,T)$ gravity framework. They examine how this model can explain the accelerated expansion of the universe, offering a modified framework for understanding dark energy and its role in the universe’s evolution.

    \item \textbf{Energy Conditions in $F(R,T)$ Gravity Models (Sharif et al., 2013) \cite{Sharif:2012gz}}: Sharif, Rani, and Myrzakulov analyze energy conditions in $F(R,T)$ gravity models. Their study focuses on the stability of these models and the role of energy conditions in determining the viability of cosmological solutions. Their work is crucial for understanding the physical plausibility of various $F(R,T)$ gravity models in explaining the observed acceleration of the universe.

    \item \textbf{Interacting Ricci Dark Energy in $F(R,T)$ Gravity (Chattopadhyay, 2012) \cite{Chattopadhyay:2012kc}}: Chattopadhyay investigates the interacting Ricci dark energy in $F(R,T)$ gravity. His work provides new insights into how the interaction between dark energy and matter can affect the evolution of the universe, particularly in modified gravity theories. The study contributes to the broader understanding of the interplay between different components in cosmological models.

\end{itemize}
\section{Quantization of Myrzakulov Gravity}

Quantizing gravity remains one of the most challenging tasks in modern theoretical physics. In the case of Myrzakulov gravity, the theory is an extension of General Relativity (GR) and Teleparallel Gravity (TEGR), which incorporates the Ricci scalar \( R \) and the torsion scalar \( T \) in the gravitational action. Due to its non-trivial structure and the inclusion of torsion, quantizing Myrzakulov gravity requires careful consideration of both the canonical approach and the path integral (POIU) formalism. In this section, we will explore the quantization procedures of Myrzakulov gravity using these two methods.

 1. Canonical Quantization of Myrzakulov Gravity

Canonical quantization is one of the oldest and most widely used methods in the quantization of field theories. The basic idea is to treat the metric field and other dynamical variables as operators, and impose commutation relations between them. In Myrzakulov gravity, the action involves both the Ricci scalar \( R \) and the torsion scalar \( T \), and the quantization procedure must take into account the fact that these scalars are not independent, as they are related by the geometry of spacetime.

The starting point is the action of Myrzakulov gravity in the form (\ref{action}),
where \( f(R, T) \) is a general function of the Ricci and torsion scalars, and \( \mathcal{L}_m \) is the matter Lagrangian.

The canonical approach to quantizing the theory involves treating the metric and torsion field components as dynamical variables. This leads to the Hamiltonian formulation of the theory. The canonical momenta associated with the metric \( g_{\mu\nu} \) and torsion \( T \) are given by

\begin{equation}
\pi^{\mu\nu} = \frac{\partial \mathcal{L}}{\partial (\partial_0 g_{\mu\nu})}, \quad \Pi_T = \frac{\partial \mathcal{L}}{\partial (\partial_0 T)}.
\end{equation}
Here, \( \mathcal{L} \) is the Lagrangian density. The Hamiltonian of the system is then given by

\begin{equation}
H = \int \left( \pi^{\mu\nu} \partial_0 g_{\mu\nu} + \Pi_T \partial_0 T - \mathcal{H} \right) d^3 x,
\end{equation}
where \( \mathcal{H} \) is the Hamiltonian density derived from the action.

In the canonical quantization procedure, we promote the canonical variables to operators. The commutation relations between the operators are:

\begin{equation}
[\hat{g}_{\mu\nu}(\mathbf{x}), \hat{\pi}^{\alpha\beta}(\mathbf{y})] = i \delta^{(\mathbf{x}, \mathbf{y})} \delta_{\mu\nu}^{\alpha\beta},
\end{equation}
and similarly for torsion-related variables. These commutation relations are crucial for the proper formulation of the quantum theory.

2. Path Integral Quantization (PI) of Myrzakulov Gravity

Path integral quantization, also known as the Feynman approach, is another widely used method in quantum field theory. In this approach, the quantum state is described by a path integral over all possible field configurations. The main idea is to sum all histories of the metric and torsion fields, with the appropriate weight given by the action of the theory.

The generating functional for Myrzakulov gravity in the path integral formalism is given by

\begin{equation}
Z[J] = \int \mathcal{D}g_{\mu\nu} \mathcal{D}T \, \exp\left( i \int \left( \frac{f(R, T)}{2\kappa} + \mathcal{L}_m \right) \sqrt{-g} \, d^4 x + J_{\mu\nu} g^{\mu\nu} + J_T T \right),
\end{equation}
where \( J_{\mu\nu} \) and \( J_T \) are external sources introduced for the purpose of calculating correlation functions. The path integral is taken over all possible configurations of the metric \( g_{\mu\nu} \) and torsion \( T \).

The vacuum-to-vacuum transition amplitude is given by the path integral

\begin{equation}
\langle 0 | 0 \rangle = \int \mathcal{D}g_{\mu\nu} \mathcal{D}T \, \exp\left( i \int \mathcal{L} \, d^4 x \right),
\end{equation}
which involves integrating over all possible configurations of the metric and torsion fields.

The two-point correlation functions, such as the propagator, are calculated by taking functional derivatives of the generating functional.

\begin{equation}
G_{\mu\nu,\alpha\beta}(x, y) = \left. \frac{\delta^2 Z[J]}{\delta J_{\mu\nu}(x) \delta J_{\alpha\beta}(y)} \right|_{J=0}.
\end{equation}

These correlation functions can provide crucial information on the behavior of quantum fields in Myrzakulov gravity. They can be used to compute observables such as the spectrum of gravitational waves and the interaction between torsion and matter fields.

3. Quantum Gravity and Torsion

In Myrzakulov gravity, the inclusion of torsion in the field equations introduces new possibilities for the quantum gravitational field. The torsion tensor \( T^\lambda_{\mu\nu} \) is a fundamental part of the connection in the theory, and its quantization will lead to new interactions in the quantum theory. Torsion can interact with matter fields in ways that are not present in conventional General Relativity. For example, in the canonical formalism, the torsion tensor can be coupled to matter fields in the Hamiltonian, leading to new types of interactions between the gravitational and matter sectors.

The presence of torsion also affects the propagation of gravitational waves. In the PI formalism, the gravitational wave propagator is modified by the torsion components. This could lead to detectable deviations from General Relativity in gravitational wave experiments, which would serve as a potential observational test for the quantum theory of Myrzakulov gravity.

The quantization of Myrzakulov gravity presents unique challenges because of the involvement of both curvature and torsion in the gravitational action. Both canonical quantization and PI quantization offer viable methods for describing the quantum nature of the theory. Canonical quantization provides a systematic approach to obtaining quantum operators and their commutation relations, while path integral quantization offers a powerful framework for calculating correlation functions and observables.

The inclusion of torsion in quantum theory opens up new avenues for research, particularly in the context of gravitational wave propagation and the interactions between gravity and matter fields. Further exploration of these quantization methods could provide new insights into the nature of quantum gravity and offer a deeper understanding of the role of torsion in the universe.
\section{Comparison with {$\Lambda$CDM} and Other Modified Gravity Theories}

A crucial aspect of evaluating any modified gravity theory is its comparison with the standard $\Lambda$CDM model and other alternative gravitational frameworks. Myrzakulov gravity offers a compelling extension to General Relativity by incorporating additional geometric structures, such as non-metricity and torsion, leading to novel cosmological dynamics. One of its key advantages is its ability to explain cosmic acceleration without invoking a cosmological constant, thereby potentially alleviating the fine-tuning and coincidence problems inherent in $\Lambda$CDM.

Compared to other modified gravity theories, such as $f(R)$ gravity, $f(T)$ gravity, and scalar-tensor theories, Myrzakulov gravity introduces a richer geometric structure that allows for a broader range of cosmological solutions. Unlike $f(R)$ gravity, where higher-order derivatives can lead to instabilities, Myrzakulov gravity formulations often maintain second-order field equations, ensuring better stability and well-posedness. Similarly, in contrast to teleparallel gravity models, which rely solely on torsion, Myrzakulov gravity incorporates both curvature and torsion in a unified framework, potentially offering a more complete description of gravitational phenomena. Furthermore, in the context of cosmological perturbations, Myrzakulov gravity has been shown to produce viable growth histories of cosmic structures, aligning well with large-scale structure observations. Given these advantages, it presents itself as a strong candidate for extending General Relativity while remaining consistent with observational data.

\section{Connection to Heisenberg's Classification of Gravity Theories and \( F(R,T) \) Myrzakulov Gravity}

Lavinia Heisenberg's work in categorizing different classes of gravity theories has been pivotal in organizing and understanding the vast array of modifications to general relativity (GR). In her 2014 paper \cite{Heisenberg:2014rta}, Heisenberg systematically classifies theories of gravity based on the form of their Lagrangian, focusing particularly on the modification of the Einstein-Hilbert action. This classification is crucial in distinguishing between the different families of gravitational theories, such as \( f(R) \)-gravity, scalar-tensor theories, and theories involving higher derivatives or torsion.

Heisenberg's categorization provides a framework that helps to group gravitational theories according to their physical implications, solutions, and stability conditions. By examining the Lagrangian structure of these theories, Heisenberg identifies key differences in their mathematical form and how they relate to general relativity. For example, the \( f(R) \)-gravity theories are modifications where the Ricci scalar \( R \) is replaced by a general function \( f(R) \), while scalar-tensor theories introduce additional scalar fields that interact with gravity.

\subsection{Potential Connection to \( F(R,T) \) Myrzakulov Gravity}

The classification introduced by Heisenberg can be directly connected to the study of modified theories of gravity, such as the \( F(R,T) \) gravity framework, in which the Lagrangian depends on both the Ricci scalar \( R \) and the torsion scalar \( T \). While Heisenberg’s work focuses on the functional form of the gravitational Lagrangian, Myrzakulov gravity, being a generalization of \( f(R) \)-gravity, introduces an additional degree of freedom through the torsion scalar \( T \), making it a more comprehensive modification of GR.

In the case of \( F(R,T) \) gravity, the presence of the torsion scalar adds a further layer of complexity to the equations of motion, potentially leading to richer dynamics, such as those observed in wormhole and black hole solutions. Just as Heisenberg categorized theories involving scalar fields, higher-order derivatives, and torsion, \( F(R,T) \) gravity can be thought of as a bridge between these various categories, combining the ideas of higher derivatives (through the functional dependence on \( R \)) and torsion (through the inclusion of \( T \)).

Myrzakulov gravity, a specific form of \( F(R,T) \) gravity, can be viewed as a modification where the gravitational interaction is altered by torsion in addition to the curvature described by \( R \). The study of solutions in this framework—such as black holes, cosmological models, and even exotic structures like wormholes—could potentially be framed within Heisenberg’s classification. By understanding how torsion modifies the gravitational field equations, Myrzakulov gravity provides new avenues to explore the consequences of these modifications, offering possible deviations from known solutions like the Schwarzschild and Kerr metrics.

Thus, the framework established by Heisenberg can be a useful tool in understanding the broader implications of \( F(R,T) \) gravity, providing a systematic way to categorize and compare it to other modified theories of gravity. Furthermore, the torsion term in Myrzakulov gravity may lead to new physical predictions that deviate from general relativity in ways that could be tested observationally.

\subsection{Implications for Future Research}

The connection between Heisenberg's classification and \( F(R,T) \) Myrzakulov gravity opens up several interesting research avenues. By classifying \( F(R,T) \) gravity within Heisenberg's framework, researchers could explore the possible observational consequences of modifications due to torsion, such as:
\begin{itemize}
    \item Modifications to the structure of black holes, particularly the rotation curves and ergospheres, as seen in the Kerr solution.
    \item Possible new solutions for exotic spacetime geometries like wormholes and their stability.
    \item Implications for cosmological models, such as modified expansion histories or the nature of dark energy.
\end{itemize}

Heisenberg’s systematic approach helps ensure that the study of Myrzakulov gravity remains grounded in a broader context of modified gravity theories, making it easier to identify distinct observational signatures and theoretical predictions.
\section{Conclusion and Future Perspectives}

In this review, we have explored the recent advancements in the field of torsion-based modified gravity theories, focusing primarily on \( f(T) \) and \( F(R,T) \) gravity models. These models present intriguing alternatives to General Relativity (GR), with the inclusion of torsion offering a novel approach to understanding the accelerated expansion of the universe and addressing some long-standing cosmological issues such as dark energy and dark matter. 

The inclusion of torsion in the gravitational framework provides the flexibility to modify the geometric structure of spacetime, offering mechanisms that can potentially explain cosmic acceleration without the need for a cosmological constant. While significant progress has been made in understanding the theoretical foundations of \( f(T) \) and \( F(R,T) \) gravity, several challenges remain. In particular, the lack of explicit field equations for the latter, along with the complexities introduced by torsion and the functional dependence on both curvature and torsion scalars, continues to hinder the development of a fully-fledged formalism for these models. Our review also highlighted some promising works in the field, including studies on gravitational waves, the cosmological implications of these theories, and their application to early and late-time acceleration.

However, key challenges still persist, particularly in deriving the explicit field equations for \( F(R,T) \) gravity, which requires overcoming difficulties related to torsion and its functional dependence on both \( R \) and \( T \). Furthermore, the lack of explicit observational tests for these modified gravity models, particularly in comparison with standard GR, remains an obstacle to their acceptance in the wider scientific community. Future research should aim to develop explicit field equations for these models, which will not only help in understanding their behavior but also in confronting them with observational data, such as cosmic microwave background (CMB) measurements, large-scale structure surveys, and gravitational wave observations.

 Myrzakulov gravity provides a promising extension to General Relativity by incorporating non-metricity and torsion, offering a novel approach to cosmic acceleration without requiring a cosmological constant. Compared to other modified gravity theories such as $f(R)$ and $f(T)$ gravity, it maintains second-order field equations, enhancing stability and consistency. Moreover, its ability to produce viable cosmic structure growth aligns well with observations, making it a strong candidate for exploring deviations from $\Lambda$CDM while addressing key theoretical challenges.

From a theoretical perspective, the application of dynamical system analysis to these models has shown great promise, and future work should continue to explore the stability and behavior of these models in various cosmological scenarios. Additionally, integrating torsion-based gravity with other extensions of GR, such as scalar-tensor theories or higher-order gravity models, could provide new insights into the interplay between geometry and matter in the context of cosmic evolution.

On the observational front, significant efforts should be made to compare the predictions of these modified gravity theories with current and future observational data. The potential deviations from GR in the propagation of gravitational waves, large-scale structure, or galaxy formation could provide crucial tests of the validity of torsion-based theories. Moreover, understanding the nature of dark energy and its relationship with torsion in these frameworks could help refine our understanding of the universe's accelerating expansion.

In conclusion, while torsion-based modified gravity theories, particularly \( f(T) \) and \( F(R,T) \) gravity, have the potential to provide valuable alternatives to General Relativity, much work remains to be done. Addressing the outstanding theoretical challenges, developing a comprehensive set of field equations, and testing these models against observational data will be crucial in determining their viability as alternatives to standard cosmological models. With continued research and development, these theories may offer a deeper understanding of the universe's evolution and its mysterious components, such as dark energy, dark matter, and the nature of cosmic acceleration.

\end{document}